\documentclass[aps,pra,twocolumn,floatfix]{revtex4-2}
\usepackage{amsmath,epsfig,mathrsfs,graphicx,amssymb,color,float}
\usepackage[T1]{fontenc}
\usepackage{t1enc}
\usepackage{hyperref}
\usepackage{array}
\usepackage{booktabs}
\usepackage{ulem}

\def\be#1\ee{\begin{equation}#1\end{equation}}
\newcommand{\ba}{\begin{eqnarray} }
	\newcommand{\ea}{\end{eqnarray} }
\def\cor#1{{#1}}

\usepackage{amsmath}            
\usepackage{amsfonts}           
\usepackage{amssymb}            
\usepackage{enumerate,graphicx}
\usepackage{array}
\usepackage{booktabs}
\usepackage{bbm}
\usepackage{physics}

\usepackage{amsmath}            
\usepackage{amsfonts}           
\usepackage{amssymb}            
\usepackage{enumerate,graphicx}
\usepackage{array}
\usepackage{booktabs}
\usepackage{yquant}
\usetikzlibrary{fit, quotes}
\usetikzlibrary{snakes,arrows,shapes}
\useyquantlanguage{groups}
\usepackage{physics}
\usepackage{todonotes}
\newtheorem{thm}{Theorem}

\def\mb{\begin{pmatrix}}
	\def\me{\end{pmatrix}}
\def\be#1\ee{\begin{equation}#1\end{equation}}

\def\mb{\begin{pmatrix}}
	\def\me{\end{pmatrix}}
\def\be#1\ee{\begin{equation}#1\end{equation}}

\begin{document}
	\title{Testing time order and Leggett-Garg inequalities with noninvasive measurements
		on public quantum computers}

	\author{Tomasz Rybotycki$^{1,2,3}$}
	\author{Tomasz Bia{\l}ecki$^{4,5}$}
	\author{Josep Batle$^{6,7}$}
	\author{Bart{\l}omiej Zglinicki$^4$}
	\author{Adam Szereszewski$^4$}
	\author{Wolfgang Belzig$^8$}
	\author{Adam Bednorz$^4$ (corresponding author)}
	\email{abednorz@fuw.edu.pl}
	
	\affiliation{$^1$Systems Research Institute, Polish Academy of Sciences, ul. Newelska 6,
		01-447 Warsaw, Poland}
	\affiliation{$^2$Nicolaus Copernicus Astronomical Center, Polish Academy of Sciences,
		ul. Bartycka 18, 00-716 Warsaw, Poland
	}
	\affiliation{$^3$Center of Excellence in Artificial Intelligence, AGH University,
		al. Mickiewicza 30, 30-059 Cracow, Poland
	}
	\affiliation{$^4$Faculty of Physics, University of Warsaw, ul. Pasteura 5, PL02-093
		Warsaw, Poland}
	\affiliation{$^5$Faculty of Physics and Applied Informatics, University of Lodz,
		ul. Pomorska 149/153, PL90-236 Lodz, Poland}
	\affiliation{$^6$Departament de F\'isica and Institut d'Aplicacions Computacionals
		de Codi Comunitari (IAC3), Campus UIB, E-07122 Palma de Mallorca, Balearic Islands,
		Spain}
	\affiliation{$^7$CRISP -- Centre de Recerca Independent de sa Pobla, 07420 sa Pobla,
		Balearic Islands, Spain}
	\affiliation{$^8$Fachbereich Physik, Universit{\"a}t Konstanz, D-78457 Konstanz, Germany}

	\begin{abstract}

We demonstrate the first violation of the Leggett-Garg inequality and time-order noninvariance on public quantum computers using genuine noninvasive measurements.	
 By gathering sufficiently large statistics, we have been
		able to violate Leggett-Garg inequality and time-order invariance. The detailed
		analysis of the data on 10 qubit sets from 5 devices available on IBM Quantum and one on IonQ reveals violations beyond 5
		standard deviations in almost all cases. We implemented our protocols using fractional gates, newly available on the IBM Heron devices, allowing us to benchmark them in application to weak measurements. The noninvasiveness is supported by a
		qualitative and quantitative agreement with the model of weak disturbance. 
Moreover, our data expose statistically significant deviations from theoretical predictions that exceed declared device error rates, establishing weak measurement protocols as a sensitive benchmark for quantum hardware.
These advances transform public quantum computers into practical testbeds for probing foundational questions of realism and temporal order with unprecedented accessibility and precision.
	\end{abstract}
	\maketitle
	
	\section{Introduction}
	Quantum physics defies classical intuitions, most prominently in the violation of local
	realism \cite{epr,bell,chsh}. The root of many nonclassical properties is that the
	values of the observables do not exist before measurement --- they are 
	probabilistic~\cite{mer}. The problem occurs for incompatible observables, which are
	expressed formally by noncommuting operators. Attempts to define the values of
	incompatible observables, even when not measured, always end up with a disagreement with
	some classical expectation. For example, Wigner function could be the joint probability
	of incompatible position and momentum, but it is negative in specific
	cases~\cite{wigner}. On the other hand, one cannot constantly apply a projective measurement
	as it freezes systems' dynamics (so called quantum Zeno effect \cite{zeno}). An incompatible projective
	measurement disturbs further evolution, e.g. in Elitzur-Vaidman bomb~\cite{elva,elvax}.
	Another long-standing conflict with quantum reality is violation of Leggett-Garg (LG)
	inequality~\cite{lega,emary}, under the assumption of noninvasiveness and macrorealism.
	In its original from, LG inequality shows that strong, projective measurements \cite{neumann}
	violate these assumptions. This claim was shown in numerous experiments
	\cite{lginv1,lginv2}. Since projective measurements seem to be highly invasive, their
	outcomes cannot reveal realistic properties of a quantum system.
	
	A partial solution is measuring not projectively, but with a finite strength. The
	smaller the strength, the better. It can be done by adding an interaction with an
	auxiliary detection space, equivalent to a positive operator-valued measure
	(POVM)~\cite{kraus,povm,peres,nielsen}. Such a measurement leads to a balance between
	accuracy and disturbance. In time-continuous version, the measurement causes a collapse,
	becoming effectively a projection, within the timescale depending on the measurement
	strength~\cite{grw,cont}. The POVM is noninvasive only if it is commuting with all the relevant
	observables, e.g. identity~\cite{busch}. 
	
\begin{figure}
		\includegraphics[scale=.3]{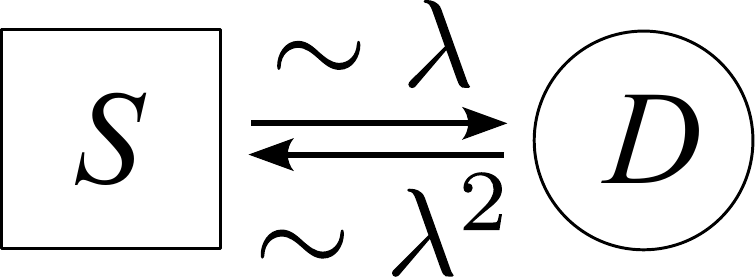}
		\caption{The scheme of weak measurement. The system $S$ interacts with the ancillary
			detector $D$ which gains information from the weakly-measured system scaled by the measurement
			strength $\lambda$, while causing disturbance to the system of the
			order $\lambda^2$.}
		\label{weakscheme}
	\end{figure}

	The strongest candidate for a genuine noninvasive measurement is the so-called weak
	measurement. It is modeled so that it has small strength and disturbance~\cite{aav}. 
	In the weak measurement scheme, the auxiliary detector gains information from the weakly-measured system
	in a form of an additional signal on top of the device noise, scaled by the measurement
	strength. The latter is meant to be small. Notice that in this scenario the disturbance
	of the weakly measured system is also weak, as it scales with the square of the
	measurement strength, see Fig. \ref{weakscheme}.	
	
	Weak measurements remain in conflict with restrictions following from classical
	realism. Their averages exceed the projective
	limits~\cite{weakx91,weakx,aweak,aweak2,lundeen}.
\cor{They allow in theory to violate LG inequalities by a
	combination of correlations in
	time~\cite{lgw1,jor1,jor2,semi3}. The experiments achieved the violation with only partial realization of weak measurements~\cite{palacios-laloy:10,gog,semi1,semi2,semi4}. In particular \cite{palacios-laloy:10} makes the measurement in the frequency domain  and
with the time correlation retrieved by the inverse Fourier transform. In \cite{lgw2,semi1,semi2} the weak measurement is mapped onto the positional degree of freedom, and in \cite{gog,semi4} one couples the detector and the system by the operation which is strong (maximally entangling),  but its action is actually weak because of specially prepared state. The optical experiments \cite{gog,lgw2,semi1,semi2} rely also on postselection on the registered photons.}
	The natural model of weak measurement~\cite{bednorz:10,abn} leads to correlations
	depending on the time order of measurements~\cite{bfb}, and violation of conservation
	of energy and other classical invariants of motion in third order
	correlations~\cite{cons}. The former issue has also been shown
	experimentally~\cite{curic} \cor{which also maps the weak measurement on the position and needs postselection}.
	
	When the public quantum computers became available via IBM Quantum platform~\cite{ibm}, 
	experiments involving LG tests have been performed, although only in the version with
	projective measurements~\cite{ibmlg1,ibmlg2}, \cor{which cannot be called noninvasive in the quantum sense.} IBM machines have also been shown to
	handle weak measurements~\cite{ibmweak}. In the hereby paper, we demonstrate violation
	of LG inequalities and time order symmetry using weak measurements. \cor{In contrast to previous experiments we \begin{itemize}
\item construct the implementation on a public quantum computer, making the test easily reproducible
\item avoid postselection
 by using stable ion or transmon qubits,
\item implement the weak detectors clearly separated from the system by ancilla qubits
\item weakly couple the system and ancilla independently of the initial detector state by using fractional gates
\item apply two weak measurements instead of one to run the test of time order violation simultaneously with LG violation.
\end{itemize}
The last task is challenging in the sense of large statistics to be collected to reduce errors, i.e. each weak measurement increases the uncertainty by its inverse strength.}

We run our
	experiments on publicly available quantum devices from IBM and IonQ. For the former, we
	used 4 different devices. We started with the simulations, during which we identified ten 3-qubit groups that were most likely to violate the LG inequalities. Each simulation was run with the noise model of a respective
	device. We then applied two protocols of weak measurements, one with a single \cor{(as in \cite{gog,semi4})} and the
	latter with two standard two-qubit gates, including new fractional gates available on IBM Heron generation devices~\cite{fgates}. The results confirm LG and time-symmetry
	violation on most of the groups by 10 standard deviations on average. We have also
	checked that the measurement is indeed noninvasive, i.e. the correlations are different
	from the ones obtained by applying projective measurements. In the case of IonQ we have selected a single 3-qubit group on a single device. We also used a single entangling gate. Our choice was
	made due to all-to-all connectivity on the selected device.
	
	The paper is organized as follows. First, we summarize the concept of weak measurements,
	LG inequalities and time order. Next, we describe the applied experimental protocol,
	using native qubit topology and gates of the IBM devices. Then, we present the results,
	commenting their most significant features. We finish with a discussion, stressing the
	pros and cons of our experiment. We also suggest further routes to test noninvasiveness
	and quantum realism even more convincingly.

	\section{Weak measurement}

	Suppose we have a generic POVM represented by Kraus operators $K(a)$ where $a$ is the
	measurement outcome~\cite{kraus,povm,peres,nielsen}. Then the state $\rho$ becomes
	\be
	\rho(a)=K(a)\rho K^\dag(a)\label{rab}
	\ee
	with probability $p(a)=\mathrm{Tr}\rho(a)$, and normalized by $\int da\; p(a)=1$.
	In the case of null, completely noninvasive measurement, we expect an unchanged $\rho$, i.e.
	\be
	\int da\;\rho(a)=\rho
	\ee
	A straightforward observation is that the only possibility to achieve it occurs when
	$K(a)=k(a)\mathbbm 1$ (scaled identity) for some a function $k(a)$, which is
	commonly referred to as no information gain without disturbance~\cite{busch}.
	We shall denote classical and quantum averages $\langle f(a)\rangle\equiv \int da\;
	p(a) f(a)$ and $\langle F\rangle=\mathrm{Tr}\rho F$, respectively.
	In the completely noninvasive case, the detector measures its own noise. Without loss of
	generality we assume that this noise is centered at $0$ and has a definite finite
	variance $\langle a^2\rangle$. We further assume that the information obtained from the
	system is encoded in the average $\langle a\rangle$. Then the following theorem, 
	linking quantum state discrimination and disturbance, holds.

	\begin{thm}
		If there exist two states $\rho_\pm$ in some Hilbert subspace $\mathcal H$ such that $\Delta = \langle a \rangle_{\rho_+} - \langle a \rangle_{\rho_-}$, then
		there exists such a pure state $\rho\in\mathcal H$ that 
		\be
		1-\mathrm{Tr}\rho'\rho\geq \Delta^2/4\sigma^2,
		\ee
		where
		$\rho'=\int da\; K(a)\rho K^\dag(a)$ is the output state and $\sigma^2$ is the upper bound on the variance $\langle (a-\langle a\rangle)^2\rangle$ in $\mathcal H$.
	\end{thm}

	The proof is in Methods \ref{appa}.
	The meaning of $\Delta$ is the ability of the measurement to discriminate between the states $\rho_\pm$ due to the difference of the average of $a$.
	For pure states, $\rho$ and $\rho'$, $\mathrm{Tr}\rho'\rho$ is equal to the fidelity $F(\rho',\rho)$ defined in general as $\sqrt{F}=\mathrm{Tr}\sqrt{\sqrt{\rho}\rho'\sqrt{\rho}}$ \cite{nielsen},
	and giving the lower bound on the Bures distance 
	\cite{bur} $B^2=2(1-\sqrt{F})\geq 1-F$, the trace distance $D(\rho,\rho')=\mathrm{Tr}|\rho-\rho'|/2\geq 1-\sqrt{F}$,
	and Hilbert-Schmidt distance $\mathrm{Tr}(\rho-\rho')^2\geq (1-F)^2$. 
	The restriction on the subspace $\mathcal H$ has two goals: (i) to ignore extremal
	states with artificially large $\Delta$, (ii) to allow superpositions that can be
	otherwise forbidden e.g. by superselection rules~\cite{susel}. The expression by
	fidelity allows	experimentally determined discrimination, e.g. by measuring projection
	$B=\rho$. Applying the theorem to the case $a=\pm 1$, i.e. a measurement scheme with
	dichotomic discrimination, one can simply take $\sigma^2=1$. Identifying the average
	with an observable
	\be
	A=\int da\; aK^\dag(a)K(a)
	\ee
	is attractive (as $\langle A\rangle=\langle a\rangle$), but misleading. 
	For instance, we do not necessarily have a measurement scheme to retrieve e.g. $\langle A^2\rangle$.
	To this end, we should introduce a special family of measurements $K(a)=k(a,A)$ for
	a Hermitian operator $A$ satisfying
	\be
	\int da\; |k(a,A)|^2=1.
	\ee
	For instance, a Gaussian measurement
	\be
	k(a,A)=(2\pi)^{-1/4}\exp(-(a-A)^2/4)
	\ee
	has the property $\sigma^2=1$ and 
	\be
	\langle a^2\rangle=\langle A^2\rangle+1
	\ee
	and in general the measurement probability is a convolution
	\be
	p(a)=\int da'\:n(a-a')\langle \delta(a-A)\rangle
	\ee
	with the normal distribution $n(a)=(2\pi)^{-1/2}\exp(-a^2/2)$.
	The disturbance on a generic state $\rho$ can be then described by
	\be
	\rho'=\int da\;K(a)\rho K^\dag(a)=\exp (-\mathcal G_A/8) \rho \label{eq:gdisturb}
	\ee
	where $\mathcal G_A\rho=[A,[A,\rho]]$ imposes decoherence -- reduction of the off-diagonal elements of $\rho$ in the eigenbasis of $A$.
	In particular, if $A=\int da\; |a\rangle a\langle a|$ for the orthonormal basis of states $|a\rangle$ then
	\be
	\langle \bar{a}|\rho'|a\rangle=e^{-(a-\bar{a})^2/4} \langle \bar{a}|\rho'|a\rangle.
	\ee 
	For a dichotomic scheme, with $a=\pm 1$ and 
	\be
	K(a)=\sqrt{(1+aA)/2} \label{dich}
	\ee
	restricting $|A|\leq 1$, we get disturbance 
	\be 
	2\rho'=\sqrt{1+A}\rho\sqrt{1+A}+\sqrt{1-A}\rho\sqrt{1-A}.
	\ee
	
	To get the variable-strength measurement we replace $A\to \lambda A$ with the constant
	parameter $\lambda>0$. For $\lambda\to \infty$ in the Gaussian scheme ($\lambda\to 1$ in
	the dichotomic scheme with $A^2=1$) we obtain a strong, projective measurement, while
	$\lambda\to 0$ is the limit of weak measurement. As we have seen, there  always is a
	disturbance. However, with introduction of $\lambda$, we get $\sim -\lambda^2
	\mathcal G_A/8$ in (\ref{eq:gdisturb}), in agreement with our theorem. Hence,
	small $\lambda$ justifies invasiveness. From the experimental point of view, there is no
	guarantee that the disturbance is small in the whole Hilbert space, including the
	apparently non-accessible part of the space. We have to trust the measurement scheme. This trust
	would be necessary in the classical picture, but there would not be a lower bound on
	the disturbance.
	
	We emphasize that only the limiting behavior of weak measurements is important.
	To construct their general family, we shall look into the construction of POVM,
	expressing Kraus operators by elements of a system-meter unitary transition $U$
	in the space $\mathcal H\otimes \mathcal H_M$ \cite{kraus,povm,nielsen}, 
	\be
	K(a)=\langle a|U|\Omega\rangle\label{kra}
	\ee
	with the meter states $|\Omega\rangle$ and $|a\rangle$, assuming the meter initially in
	pure $|\Omega\rangle$ state, and measuring projectively $P_a=|a\rangle\langle a|$,
	calibrated to zero average bias for null detection,
	\be
	\langle a\rangle_\Omega=\int da\; |\langle a|\Omega\rangle|^2=0.\label{unbias}
	\ee
	The weak measurement can be defined by weakening $U\to U^\lambda$, with $\lambda \to 0$.
	It needs operational feasibility  of the
	power which is realized
	for either $U\sim 1$ (already small disturbance) or continuous time evolution, e.g. $U=\exp(-i\lambda H)$. Then
	the evolution takes 
	\be
	\rho\to \lambda [H,\rho]/i\label{hhh}
	\ee
	with the anticommutator notation $[A,B]\equiv AB-BA$,
	for $U=\exp(H/i)$ and Hermitian Hamiltonian $H$. After the measurement on the meter, 
	one is left with the generic form \cite{bednorz:10,abn,bfb,povmpp,buelte:18},
	\be
	\mathcal K\rho=\int da\;aK(a)\rho K^\dag(a)\to \lambda (\{A,\rho\}/2+i[B,\rho])\label{avv}
	\ee
	with some operators $F$ and $G$, and the anticommutator notation $\{F,G\}\equiv FG+GF$. 
	The above remains true even if one adds some randomness of $U$ and an additional
	$\lambda$-independent local transformation on the meter. For the Hamiltonian
	representation (\ref{hhh}) we get 
	\be
	A/2+iB=\int da\;\langle a|H|\Omega\rangle\langle \Omega|a\rangle/i.
	\ee
	Then $\langle a\rangle\to \lambda\langle A\rangle$ which makes $B$ inaccessible in a
	single average. However, its effect can be captured by the next measurement or
	postselection. The part $B$ is rather the influence of the detector on the system,
	which can occur also classically. In fact, classical indirect measurements have similar
	features as the quantum ones \cite{bfb}, except that they can be completely noninvasive.
	This is why we shall assume $B=0$, although some measurement schemes (e.g. to measure
	emission noise \cite{povmpp,buelte:18}) contain a special form of $B$. 
	
	From now on, we shall use the weak measurement superoperator
	\be
	\mathcal A=\{A,\cdot\}/2,
	\ee
	assuming the known weak measurement protocol with some small $\lambda$ as a scaling
	factor between the detector's outcome $a$ and the observable $A$.
	In the case of symmetric dichotomic observables and projective measurements
	$A=\bar{a}(P_++P_-)$ with $K(\pm \bar{a})=P_{\pm}$, the weak and projective measurements 
	coincide as regards correlations, i.e.
	\be
	\mathcal A\rho=\sum_a aK(a)\rho K^\dag(a) \label{eq:mislead}.
	\ee
	However, equation \ref{eq:mislead} is highly misleading, as the disturbance is
	very large and can be detected with measurement uncorrelated with $a$.

	\begin{figure}
		\includegraphics[scale=.4]{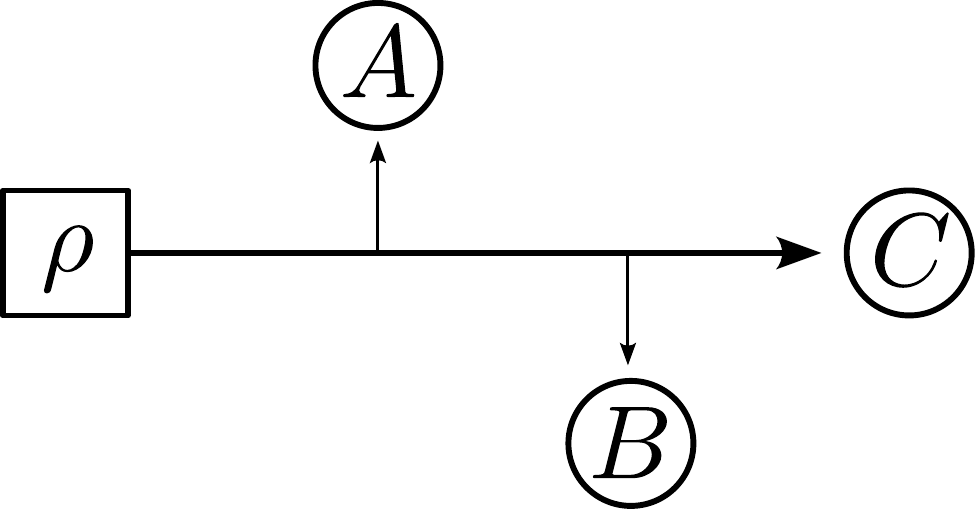}\\
		\medskip
		\includegraphics[scale=.4]{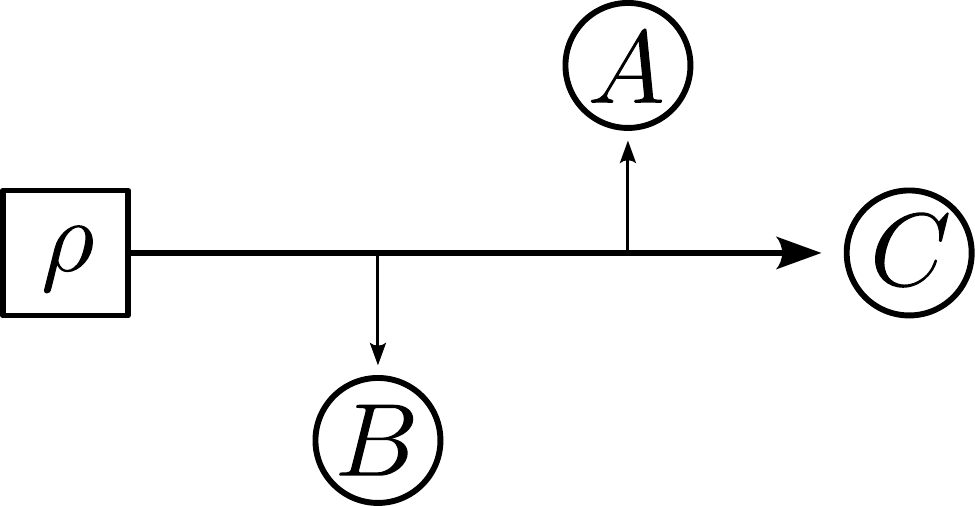}
		\caption{The detection scheme to test LG inequalities and weak order of
			measurements. Two weak detectors of $A$ and $B$ measure the state $\rho$ before the
			final measurement $C$. The time order of the measurement is from the left to the
			right: $A\to B\to C$ in the upper and $B\to A\to C$ in the lower case.}\label{lgw}
	\end{figure}

	\section{LG inequalities and time order}
	
	Suppose we have a single system and three observables, $A$, $B$, $C$ with at least one
	of them measured noninvasively, see Fig. \ref{lgw}. Then, assuming the mapping of
	the measurement outcomes, $A\to a$, $B\to b$, $C\to c$, the correlations
	$\langle a\rangle$, $\langle ab\rangle$, or $\langle abc\rangle$ reflect the properties
	of some underlying probability distribution $p(a,b,c)$. If $|a|,|b|\leq 1$ then the
	LG inequality holds
	\be
	\langle  a\rangle+\langle  b\rangle-\langle  a b\rangle\leq 1.
	\ee
	It is valid both in classical and quantum mechanics, when applying direct sequence of
	measurement with outcomes $a$ and $b$. In general, measuring two outcomes is invasive.
	The problem arises when one demands reduction of invasiveness, which is only possible in
	the limit of weak measurements. However, by our theorem, we know the invasiveness
	is connected to the smaller sensitivity and indirect interpretation of the measurement
	results $a$ and $b$. Suppose that we measure $A\to B$ (i.e. first $A$ then $B$) weakly
	(replace $a\to \mathcal A$, $b\to\mathcal B$). As we work now with superoperators, let
	us define the averages, $\langle \mathcal O\rangle\equiv\mathrm{Tr}\rho\mathcal O \mathbbm 1$
	with the superoperator $\mathcal O$ acting on identity $\mathbbm 1$.
	For the order $A\to B\to C$, where we measure weakly, we have $\langle\mathcal A\mathcal B\mathcal C\rangle=\mathrm{Tr}\rho
	\mathcal A\mathcal B\mathcal C \mathbbm 1$,
	i.e. left to right time order.
	
	Now the LG inequalities read
	\begin{align}
		&\langle\mathcal A\rangle+\langle\mathcal B\rangle-\langle\mathcal A\mathcal B\rangle\leq 1\nonumber\\
		&\langle\mathcal A\rangle+\langle\mathcal B\rangle-\langle\mathcal B\mathcal A\rangle\leq 1
		\label{lgg}
	\end{align}
	measuring $A\to B$ and $B\to A$, respectively. Moreover, classical noninvasive measurements cannot depend on their order,
	\be
	\langle \mathcal A\mathcal B\mathcal C\rangle=\langle \mathcal B\mathcal A\mathcal C\rangle \label{orr}
	\ee
	with the first order $A\to B\to C$ and the second order $B\to A\to C$. In contrast, quantum correlations, even noninvasive, depend on the order,
	\be
	\langle \mathcal A\mathcal B\mathcal C\rangle- \langle \mathcal B\mathcal A\mathcal C\rangle=
	\langle [[A,B],C]\rangle/4.
	\ee
	
	In our approach we shall (i) assume that all three observables are dichotomic $A^2=B^2=C^2=1$, (ii) the measurement scheme is also dichotomic, 
	as described by (\ref{dich}) with $A,B\to \lambda A,\lambda B$ measured weakly, and $C$ projectively (formally $\lambda_C=1$).
	The exact formulas for all correlations in this schemes are given in Methods \ref{appb}.
	
	We set initial state $\rho=|\psi\rangle\langle\psi|$ with $\sqrt{2}|\psi\rangle=|+\rangle+|-\rangle$, and the observables
	\begin{align}
		&A=e^{i\pi/4}|+\rangle\langle -|+e^{-i\pi/4}|-\rangle\langle +|\nonumber\\
		&B=e^{-i\pi/4}|+\rangle\langle -|+e^{i\pi/4}|-\rangle\langle +|\nonumber\\
		&C=i|-\rangle\langle+|-i|+\rangle\langle -|\label{pmn}
	\end{align}
	Then 
	\begin{align}
		&\langle \mathcal A\rangle=\langle\mathcal B\rangle=1/\sqrt{2}\nonumber\\
		&\langle \mathcal A\mathcal B\rangle=\langle \mathcal B\mathcal A\rangle=0,\nonumber\\
		&\langle \mathcal A\mathcal B\mathcal C\rangle=-\langle \mathcal B\mathcal A\mathcal C\rangle=1/2
	\end{align}
	which violates (\ref{lgg}) as $\sqrt{2}>1$ and (\ref{orr}) as $1/2\neq -1/2$.

	\begin{figure}
		\includegraphics{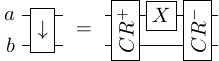}		
		\caption{The notation of the $ECR$ gate in the convention $ECR_\downarrow|ab\rangle$.}
		\label{ecr}
	\end{figure}	
	
	\begin{figure}
		\includegraphics{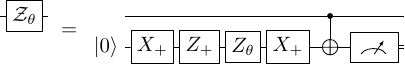}
		
		\caption{Protocol (I) with a single $CX$ gate. Weak measurement of $Z$ on the
			upper qubit, by the lower (meter) qubit with the strength of the measurement defined
			by $\sin\theta$. The control and target qubit are depicted by $\bullet$ and
			$\oplus$ respectively.}
		\label{weakc}
	\end{figure}
	
	\begin{figure}
		\includegraphics{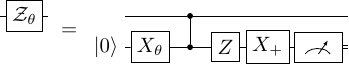}
		
		\caption{Protocol (I) with a single $CZ$ gate and a fractional $RX$, represented by $X_\theta$. Weak measurement of $Z$ on the
			upper qubit, by the lower (meter) qubit with the strength of the measurement defined
			by $\sin\theta$. The $CZ$ gate is depicted as linked $\bullet$ (the gate is
			symmetric).}
		\label{weakz}
	\end{figure}

	\begin{figure}
		\includegraphics{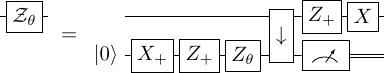}
		
		\caption{Protocol (I), with a single $ECR$ gate. Weak measurement of $Z$ on
			the upper qubit, by the lower (meter) qubit with the strength of the measurement
			defined by $\sin\theta$.}
		\label{weak1}
	\end{figure}
	
	\begin{figure}
		\includegraphics{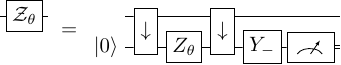}
		
		\caption{Protocol (II), with two $ECR$ gates. Weak measurement of $Z$ on the
			upper qubit, by the lower (meter) qubit with the strength of the measurement defined
			by $\sin\theta$.}
		\label{weak2}
	\end{figure}
	
	\begin{figure}
		\includegraphics{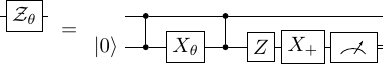}
		
		\caption{Protocol (II), with two $CZ$ gates and fractional $RX$ gate, represented by $X_\theta$. Weak measurement of $Z$ on the
			upper qubit, by the lower (meter) qubit with the strength of the measurement defined
			by $\sin\theta$.}
		\label{weak2z}
	\end{figure}

	\begin{figure}
		\includegraphics{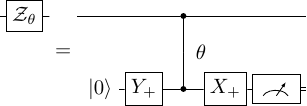}
		
		\caption{Protocol (II), with a single $(ZZ)_\theta$ gate. Weak measurement of $Z$ on
			the upper qubit, by the lower (meter) qubit with the strength of the measurement
			defined by $\sin\theta$. $(ZZ)_\theta$ gate is depicted as a link between the
			qubits.}
		\label{weakf}
	\end{figure}

	\section{Weak measurements on IBM and IonQ}

	IBM Quantum offers networks of qubits, i.e. local two-level spaces, with the default
	basis $|0\rangle$, $|1\rangle$. The states can be changed by unitary operations (gates).
	The qubit is usually measured projectively in the default basis, as the states
	differ by energy and lose mutual coherence over time. However, all projections are
	feasible, by the use of quantum gates. To realize weak measurement on a qubit, it must
	be entangled with an auxiliary meter qubit by a two-qubit operation/gate. We have
	implemented two different protocols of weak measurement:
	\begin{enumerate}
		\item[(I)] the meter is close to the special state $\sim|\Omega\rangle$ for which
		applying the operation $V'$ does not affect the system, i.e.
		$V'|\psi\rangle|\Omega\rangle=|\psi\rangle|\Omega\rangle$ for an arbitrary
		$|\psi\rangle$ \cite{wopt,wopt2,wopt3},
		
		\item[(II)] the constructed operation $U$ is $\sim 1$, i.e. close to identity
		\cite{ibmweak}.
	\end{enumerate}
	The protocol (I) involves a single entangling operation. Its small
	invasiveness relies on a specifically prepared meter state.
	On the other hand (II) technically needs either two self-reverse entangling operations
	or a single true incremental operation which is noninvasive regardless of the meter state. 
	Therefore (II) is closer to the universal derivation of weak measurements from weak
	entanglement. Nevertheless, both protocols result in the same POVM and
	their implementation on IBM is feasible. To apply (I), we can restrict $V'$ to a
	unitary $V$ in the subspace orthogonal to $|\Omega\rangle$.
	One prepares the state close to $|\Omega\rangle$,
	\be
	|\Omega_\theta\rangle=|\Omega_\perp\rangle\sin\theta/2+|\Omega\rangle\cos\theta/2,
	\ee
	where $\langle\Omega_\perp|\Omega\rangle=0$. Note that
	$\langle\Omega|V|\Omega_\perp\rangle=0$ by unitarity, so the invasiveness is
	$\sim \theta^2$. 
	We construct Kraus operators analogously to (\ref{kra}),
	\be
	K(a)=\langle a|U|\Omega_\perp\rangle\sin\theta/2+\langle a|\Omega\rangle\cos\theta/2.
	\ee
	Assuming unbiased measurement by (\ref{unbias}) and $\lambda=\sin\theta$ we get 
	the weak limits (\ref{avv}),
	\be
	A+2iB=\int da\; a\langle \Omega|a\rangle\langle a|U|\Omega_\perp\rangle.
	\ee
	In the case of purely dichotomic measurement with $a=\pm 1$ (shortly $\pm$) and
	$V^2=\mathbbm 1$ (parity operation)
	we can set 
	\begin{align}
		&\sqrt{2}|\Omega\rangle=|+\rangle+|-\rangle,\nonumber\\
		&\sqrt{2}|\Omega_\perp\rangle=|+\rangle-|-\rangle,\label{omm}
	\end{align}
	to obtain (\ref{dich}) for $A=\lambda V$.
	
	The protocol (II) also simplifies for dichotomic measurement taking
	$U_\theta=\exp(-iVW\theta/2)$ with  $V$ acting in the system space and $W$ acting in the
	meter space, such that $V^2=W^2=\mathbbm 1$, $V=V^\dag$, $W=W^\dag$. We construct the
	states (\ref{omm}) with $W|\Omega\rangle=i|\Omega_\perp\rangle$ i.e. $W$ reads in the
	basis $|+\rangle$, $|-\rangle$
	\be
	\begin{pmatrix}
		0&i\\
		-i&0
	\end{pmatrix}.
	\ee
	Then again we obtain (\ref{dich}) for $A=\lambda V$.
	
	\begin{table*}
		\begin{tabular}{*{9}{c}}
			\toprule
			G&$C$&$A$&$B$&$eCA[10^{-2}]$&$eCB[10^{-2}]$&$eC[10^{-2}]$&$eA[10^{-2}]$&$eB[10^{-2}]$\\
			brisbane&&&&&&&&\\
			0&94&95&90&1.4&0.86&1.7&2.9&0.93\\
			1&6&5&7&0.62&0.53&1.2&1&2.1\\
			2&58&71&59&0.36&0.41&1.3&1.2&1.9\\
			3&62&61&63&0.63&0.66&1.4&1.1&1.1\\
			4&52&37&56&0.58&0.8&1.3&0.9&2.3\\
			5&116&115&117&1.1&0.58&2.2&1.9&1.9\\
			6&50&49&51&1.4&1.3&2.8&0.95&3.7\\
			7&21&20&22&0.66&0.41&0.88&2.3&2.2\\
			8&125&124&126&0.68&0.79&1.4&1.4&1.6\\
			9&108&112&107&0.66&0.79&1.2&0.85&2.1\\
			\midrule
			sherbrooke&&&&&&&&\\
			0&49&50&48&0.39&1.2&5.4&3&1.6\\
			1&107&106&108&0.77&0.74&3&1.3&0.88\\
			2&119&118&120&0.67&0.76&0.78&0.9&1.1\\
			3&45&46&44&0.65&0.67&1.4&5.1&0.71\\
			4&69&68&70&0.46&0.49&2.4&1.1&17\\
			5&59&60&58&1.1&1&1.9&1.2&1.1\\
			6&89&88&74&1.1&0.82&5.4&3.7&1.2\\
			7&114&109&115&0.45&1.6&1.5&2.8&7.2\\
			8&26&25&27&0.93&0.68&0.9&2.5&0.9\\
			9&80&79&81&0.83&0.62&0.83&1.2&0.59\\
			\midrule
			kyiv&&&&&&&&\\
			0&53&60&41&0.34&0.34&0.88&0.63&0.34\\
			1&114&115&113&1.2&1.8&1.5&0.51&3\\
			2&0&1&14&0.41&0.53&0.73&1.2&1.3\\
			3&68&69&55&0.76&0.63&1.2&0.27&2.8\\
			4&26&16&25&1&1.1&2.1&3.2&1.4\\
			5&4&3&15&1&1.1&8.2&0.54&2\\
			6&103&102&104&0.48&1&0.42&2.2&1\\
			7&82&83&81&0.59&1.3&0.56&1.5&2.2\\
			8&54&45&64&0.96&1.2&0.71&0.37&0.71\\
			9&33&20&39&0.64&1.6&0.39&5.4&1.6\\
			
			\bottomrule
		\end{tabular}
		\caption{Devices from Eagle generation, groups (G) with measured qubits,  main (system) $C$,
			auxiliary (meters) $A$ and $B$, errors of $ECR/ZZ$ gates between qubit $C$ and
			$A/B$, and readout errors on qubits $C$, $A$, $B$.}
		\label{fffe}
	\end{table*}
	\begin{table*}
		\begin{tabular}{*{11}{c}}
			\toprule
			G&$C$&$A$&$B$&$CZeCA[10^{-2}]$&$RZZeCA[10^{-2}]$&$CZeCB[10^{-2}]$&$RZZeCB[10^{-2}]$&$eC[10^{-2}]$&$eA[10^{-2}]$&$eB[10^{-2}]$\\
			torino&&&&&&&&&&\\
			0&88&87&94&0.31&1.3&0.3&0.62&5.9&0.9&0.98\\
			1&29&28&36&0.33&0.62&0.22&0.39&1.5&0.85&0.56\\
			2&12&13&18&0.24&0.22&0.37&0.15&2.2&0.71&0.49\\
			3&46&47&55&0.34&0.77&0.23&0.31&4.5&1.3&0.63\\
			4&61&54&60&0.37&0.66&0.22&0.44&1.5&0.68&1.1\\
			5&33&32&37&0.3&1.3&0.28&0.26&1.3&2&0.42\\
			6&10&9&11&0.25&0.49&0.32&0.3&2.5&1.9&0.93\\
			7&80&81&92&0.25&0.45&0.15&0.3&1.6&0.81&0.39\\
			8&50&49&51&0.31&0.29&0.37&0.35&9.3&1.7&1.5\\
			9&44&35&43&0.2&0.41&0.23&0.4&1.9&1.6&0.88\\
			\midrule
			kingston&&&&&&&&&&\\
			0&129&118&128&0.12&0.14&0.17&0.085&2&0.2&0.56\\
			1&83&82&96&0.21&0.16&0.19&0.15&0.56&0.42&0.39\\
			2&141&140&142&0.15&0.14&0.14&0.16&0.49&0.63&0.61\\
			3&104&103&105&0.2&0.32&0.14&0.19&0.34&1.3&0.51\\
			4&125&117&126&0.24&0.55&0.3&0.2&7.6&0.34&0.46\\
			5&49&38&50&0.17&0.13&0.24&0.41&0.49&0.88&0.71\\
			6&14&13&15&0.14&0.3&0.6&0.18&2.4&1.1&0.63\\
			7&63&62&64&0.17&0.39&0.21&0.45&0.85&1.1&0.37\\
			8&70&69&71&0.22&0.41&0.26&0.34&1.9&0.59&0.44\\
			9&67&57&68&0.43&0.37&0.15&0.19&0.81&0.44&0.54\\
			\bottomrule
		\end{tabular}
		\caption{Devices from Heron generation, groups (G) with measured qubits,  main (system) $C$,
			auxiliary (meters) $A$ and $B$, errors of $CZ/RZZ$ gates between qubit $C$ and
			$A/B$, and readout errors on qubits $C$, $A$, $B$.}
		\label{fffh}
	\end{table*}
	
	\section{Implementation on IBM}

	For the physical implementation and manipulation of qubits as transmons \cite{transmon}
	and programming of IBM Quantum, see \cite{gam0,qurev,qisr}. To describe the actual
	implementation of the test from Sec. III, we start from Pauli matrices in the basis
	$|+\rangle\equiv |0\rangle$, $|-\rangle\equiv |1\rangle$. This is the natural
	convention in quantum computing, switching between spin and bit notation.
	The explicit form of Pauli matrices is
	\be
	X=\begin{pmatrix}
		0&1\\
		1&0\end{pmatrix},\:Y=\begin{pmatrix}
		0&-i\\
		i&0\end{pmatrix},\:Z=\begin{pmatrix}
		1&0\\
		0&-1\end{pmatrix},\:
	I=\begin{pmatrix}
		1&0\\
		0&1\end{pmatrix}.\label{pauli}
	\ee
	Now, the $|\pm\rangle$ are eigenstates of $Z|\pm\rangle=\pm|\pm\rangle$.
	The IBM Quantum devices use transmon qubits. Such systems natively support X gate
	and 
	\be
	X_+=X_{\pi/2}=(I-iX)/\sqrt{2}=\begin{pmatrix}
		1&-i\\
		-i&1\end{pmatrix}/\sqrt{2},
	\ee
	denoting $V_\theta=\exp(-i\theta V/2)=\cos(\theta/2)-iV\sin(\theta/2)$.
	Furthermore, we also denote $V_\pm= V_{\pm \pi/2}$, whenever $V^2$ is identity.
	Note that $Z_\theta=\exp(-i\theta Z/2)=\mathrm{diag}(e^{-i\theta/2},e^{i\theta/2})$ is
	a virtual gate, essentially adding the phase shift to next gates. A sequence of these
	operations allows one to realize an arbitrary unitary gate $U$. The standard projective
	measurement is implemented by $P_a=|a\rangle\langle a|$ for $a=0,1$. One can
	also see that	$Z=P_0-P_1$ represents the outcome $z=\pm 1$. The measurement can be done along any axis by using a unitary rotation $M=U^\dag Z U$. 
	
	To measure weakly, we have to couple the system qubit with an auxiliary meter. We shall
	describe two-qubit gates using the following, shorthand notation
	$(AB)|ab\rangle=(A|a\rangle)(B|b\rangle)$. The commonly used gate is controlled
	rotation, $CU$, e.g. $CX$ (also known as $CNOT$) or $CZ$. Matrix form of $CU$ gate, 
	in the $\{|00\rangle$, $|01\rangle$, $|10\rangle$, $|11\rangle\}$ basis is 
	\be
	\begin{pmatrix}
		I&0\\
		0&U
	\end{pmatrix}
	\ee
	or equivalently
	\be
	CU=(II+ZI+IU-ZU)/2.
	\ee
	For $CU|ab\rangle$ the control qubit is $a$ and the target is $b$ and $U$ is applied
	to $|b\rangle$ only if $a=1$. By local rotation $V$ on qubit $b$ one can freely
	change $U\to V^\dag UV$. In the protocol (I), the states $|\Omega\rangle$ and
	$|\Omega_\perp\rangle$ should be the eigenstates of $U$.
	For $U=X$ we have  
	\begin{align}
		&|\Omega\rangle=(|0\rangle+|1\rangle)/\sqrt{2}\nonumber\\
		&|\Omega_\perp\rangle=(|0\rangle-|1\rangle)/\sqrt{2}.
	\end{align}
	
	Protocol (I) is realized in the following way. We take
	\begin{align}
		&|\Omega_\theta\rangle=X_+Z_{\theta+\pi/2} X_+\ket 0=\nonumber\\
		&\sin(\theta/2+\pi/4)\ket 0+\cos(\theta/2+\pi/4)\ket 1,
	\end{align}
	apply $CX|\psi\Omega_\lambda\rangle$, assuming system state $\ket{\psi}$ and
	measure $Z\to\pm 1$ on the meter qubit. One can notice that we end up with the scheme
	(\ref{dich}) for $A=\lambda Z$ with $\lambda=\sin\theta$,  Fig. \ref{weakc}, i.e.
	\be
	\sqrt{2}K_\pm=\cos\theta/2+Z\sin\theta/2.
	\ee
	Instead of $CX$, the native gate on IBM Heron devices, like \texttt{ibm\_torino}, is $CZ$
	\cite{czg}. Moreover, these devices support a new single-qubit fractional $RX(\theta)\equiv X_\theta$ gate.
	This enforces a slight modification of protocol (I) implementation.
	The prepared meter state is 
	\be
	|\Omega_\theta\rangle=X_\theta\ket 0=\cos(\theta/2)\ket 0-i\sin(\theta/2)\ket 1
	\ee
	instead, and one measures $-Y$, see Fig. \ref{weakz}.

	The IBM Quantum devices use a native asymmetric two-qubit $ECR$ gates instead of $CX$,
	but one can transpile the latter with the former, by adding single-qubit gates. 
	The $ECR$ gate acts on the states $|ab\rangle$ as (Fig. \ref{ecr})
	\ba
	&ECR_\downarrow=((XI)-(YX))/\sqrt{2}=CR^- (XI) CR^+=\nonumber\\
	&
	\begin{pmatrix}
		0&X_-\\
		X_+&0\end{pmatrix}
	=\begin{pmatrix}
		0&0&1&i\\
		0&0&i&1\\
		1&-i&0&0\\
		-i&1&0&0\end{pmatrix}/\sqrt{2},
	\ea
	in the basis $|00\rangle$, $|01\rangle$, $|10\rangle$, $|11\rangle$,
	with Crossed Resonance gates
	\be
	CR^\pm=(ZX)_{\pm \pi/4}.
	\ee
	The gate is its inverse, i.e. $ECR_\downarrow ECR_\downarrow=(II)$.
	The protocol (I) can be still realized by $CX=(XI)(Z_+I)ECR_\downarrow(IX_-)$, see Fig.
	\ref{weak1}. The protocol (II) needs an operation $\sim 1$ which is realized by two
	$ECR$ gates with a small phase shift on the meter (target) qubit
	\begin{align}
		&(ZY)_\theta=II\cos(\theta/2)-iZY\sin(\theta/2)\nonumber\\
		&ECR_\downarrow(IZ_\theta)ECR_\downarrow
	\end{align}
	on IBM Eagle. The meter qubit is initially in the state
	$|0\rangle$ and finally measured by $X=\pm 1$. It is possible by applying 
	$Y_-$ rotation before readout since $Y_+ZY_-=-iYZ=X$, see Fig. \ref{weak2}. In the case of IBM Heron devices, one uses two $CZ$ gates and again a fractional $RX(\theta)$ gate, followed by measuring $-Y$, see Fig. \ref{weak2z}. The resulting POVM is identical to the protocol (I). Both protocols
	measure any unit combination of Pauli matrices by appropriate unitary transform 
	$Z\to U^\dag ZU$, with $U$ and $U^\dag$ applied before and after the protocol,
	respectively.
	
	Importantly, the protocol (II) on Heron and IonQ uses native symmetric fractional gates $(ZZ)_\theta$,
	\begin{equation}
		\begin{pmatrix}
			e^{i\theta/2}&0&0&0\\
			0&e^{-i\theta/2}&0&0\\
			0&0&e^{-i\theta/2}&0\\
			0&0&0&e^{i\theta/2}
		\end{pmatrix}
	\end{equation}
	with the initial detector's state $(|0\rangle-i|1\rangle)/\sqrt{2}=X_+|0\rangle$ and
	measuring along $X$ basis by applying $Y_-$ before the measurement. Since IBM
	allows only $\theta>0$, we reversed the sign by $Y_+ \to Y_-$, see Fig. \ref{weakf}.
	One can find further details about the circuits we used in Methods \ref{appc}.

	\section{Results}

	We have tested 10 sets of 3 qubits on 3 IBM Eagle devices: \texttt{ibm\_brisbane},
	\texttt{ibm\_sherbrooke}, \texttt{ibm\_kyiv}, and 2 IBM Heron devices \texttt{ibm\_torino} and  \texttt{ibm\_kingston}, see details in
	Tables \ref{fffe}, \ref{fffh}. We ran 60/40 jobs on Eagle/Heron, each one with 10000 shots, 25
	repetitions of $A\to B\to C$ and $B\to A\to C$ measurement orders for $\theta=\pm 0.1$
	(i.e. $2\times 2$ measurements). The actual measurement is the projection on the states
	$|abc\rangle$ where $a$, $b$ are the outcomes of auxiliary meter qubits, corresponding
	to weak measurement of $A$ and $B$, respectively, while $c$ is the final projection on
	the system qubit. To eliminate drifts, we have taken the difference of results for
	$\theta=\pm 0.1$ and divided the difference by $2\lambda=2\sin|\theta|$. It gives a
	total of 8 circuits per measurement scheme, with $25\times 8=200$ circuits in each job,
	randomly shuffled to avoid memory-related effects. 
	
\begin{figure}[ht]
		\includegraphics[scale=.5]{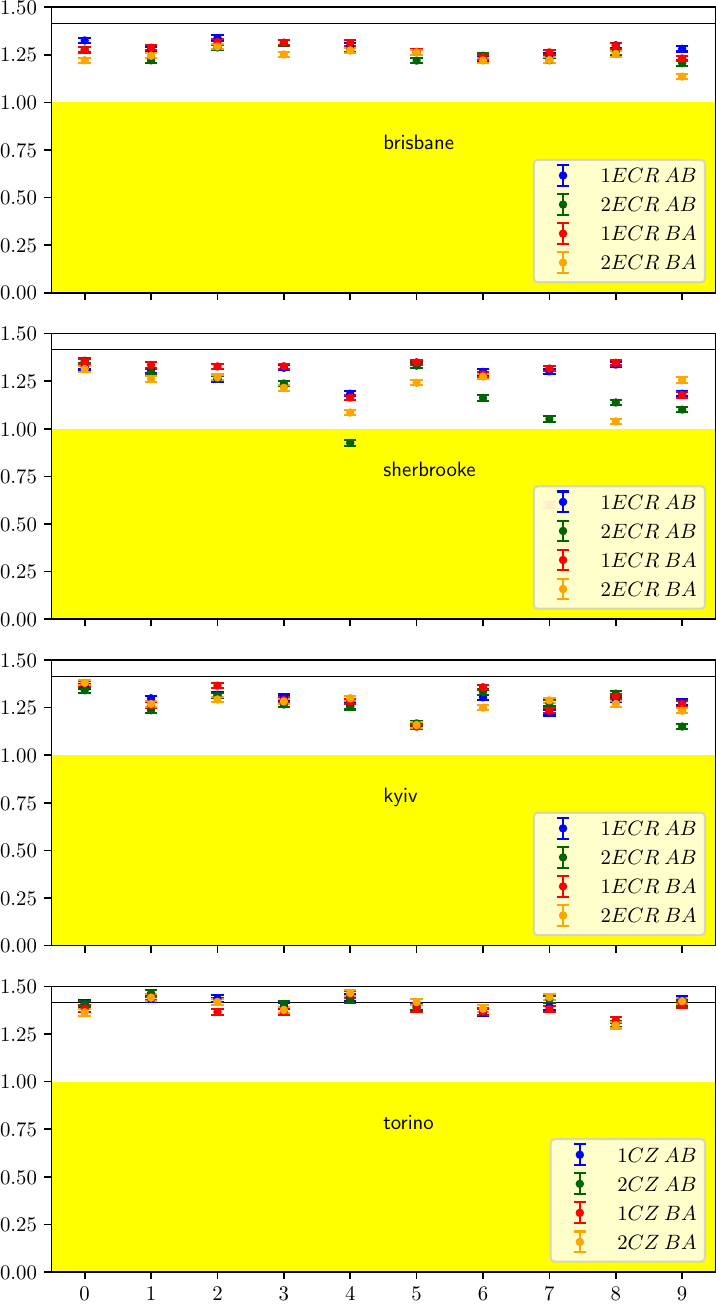}
		\caption{Leggett-Garg violation for qubits from the tested IBM devices (see Tables \ref{fffe},\ref{fffh}).
			The values correspond to the combination of correlations taken from Eq. (\ref{lgab}).
			The results $nECR/CZ$ correspond to $n=1,2$ gates $ECR/CZ$, while $AB$ and $BA$ correspond to the order $A\to B$ and $B\to A$.
			The solid line is the perfect value $\sqrt{2}$ while the yellow region is below the classical limit.
		}
		\label{lgbr}
	\end{figure}

\begin{figure}[ht]
		\includegraphics[scale=.5]{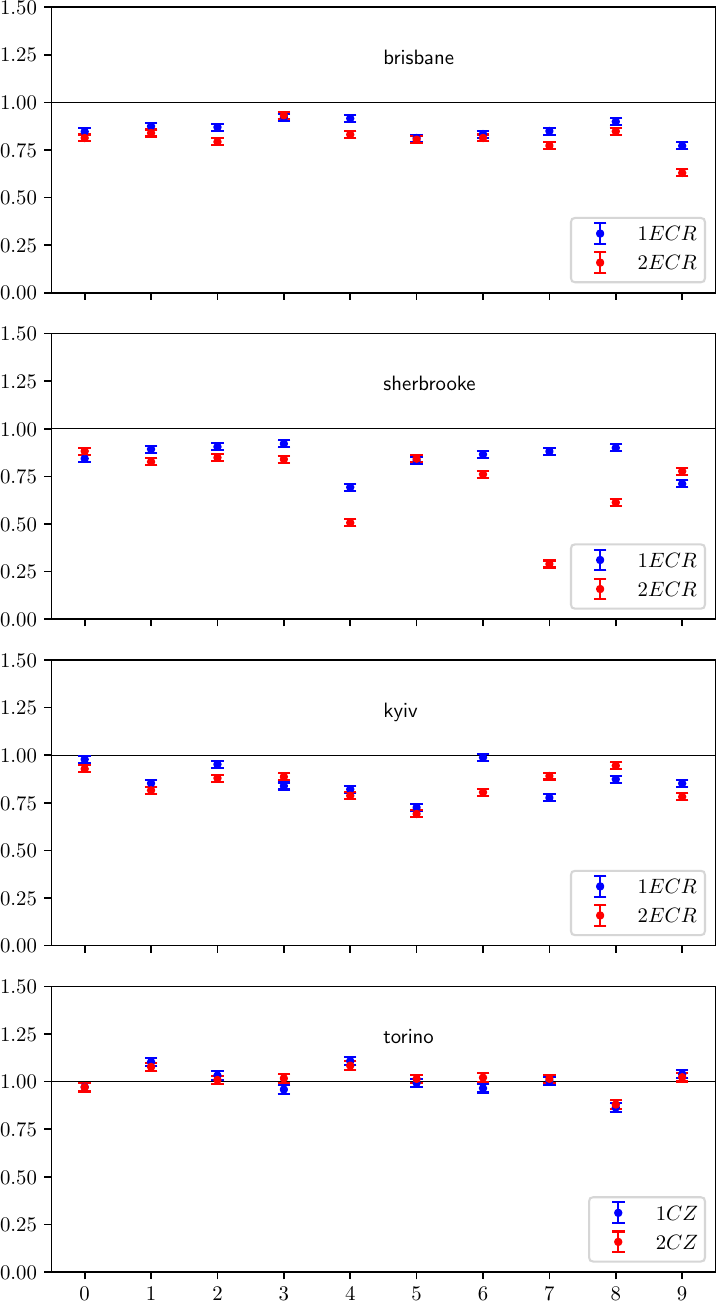}
		\caption{Order invariance violation for qubits from the tested IBM devices (see Tables \ref{fffe}, \ref{fffh}).
			The values correspond to the difference $\langle \mathcal A\mathcal B\mathcal C\rangle-\langle\mathcal B\mathcal A\mathcal C\rangle$.
			The results $nECR/CZ$ correspond to $n=1,2$ gates $ECR/CZ$
			The solid line is the perfect value $1$.
		}
		\label{ord}
	\end{figure}

	The results of Leggett-Garg test on observables $A$ and $B$, are shown in Fig.
	\ref{lgbr} while the test of time order is presented in Fig. \ref{ord}. We denoted
	weak measurement performed by protocols (I) and (II) by $1ECR/CZ$ and $2ECR/CZ$,
	respectively. To calculate the error, we assumed that shots are identical and
	independent of each other. This allowed us to use the Bernoulli formula in our analysis. For
	weak measurements, $\mathcal A$ and $\mathcal B$, the actual product of values $ab$ is
	almost uniform with random values $\pm 1$. This makes the almost identical error of each of correlations $Q$ of the form
	$\langle \mathcal A\mathcal B\rangle$
	or
	$\langle\mathcal A\mathcal B\mathcal C\rangle$
	and $\mathcal A\leftrightarrow\mathcal B$. It is just scaled by the assumed measurement strength.
	We add up all 4   strength combinations to get the final error scaled by the total number of experiments,
	\be
	\sqrt{\langle (Q-\langle Q\rangle)^2\rangle}\simeq 1/2\sqrt{JSR}\sin^2(\theta)
	\ee
	for $J$ jobs $S$ shots and $R$ repetitions, here $\simeq 0.013$. It is
	much smaller than the observed violations.
	
	In the case of LG, we summed up errors treating correlations as independent of
	single average $\langle \mathcal A\rangle$ and $\langle\mathcal B\rangle$, but the
	correlation contribution dominates at least $1/\lambda$ times. There is also a small
	correction by invasiveness, of the order $\lambda^2$, which is negligible in this case,
	see Methods \ref{appd}. The LG violation is about 10 standard deviations, except for 
	isolated sets of qubits. The same applies to the time order violation.
	Note that the third order correlation is identical when
	applying a sequence of 3 projective measurements, but then the invasiveness is
	obvious. 
	
\begin{figure}[ht]
		\includegraphics[scale=.5]{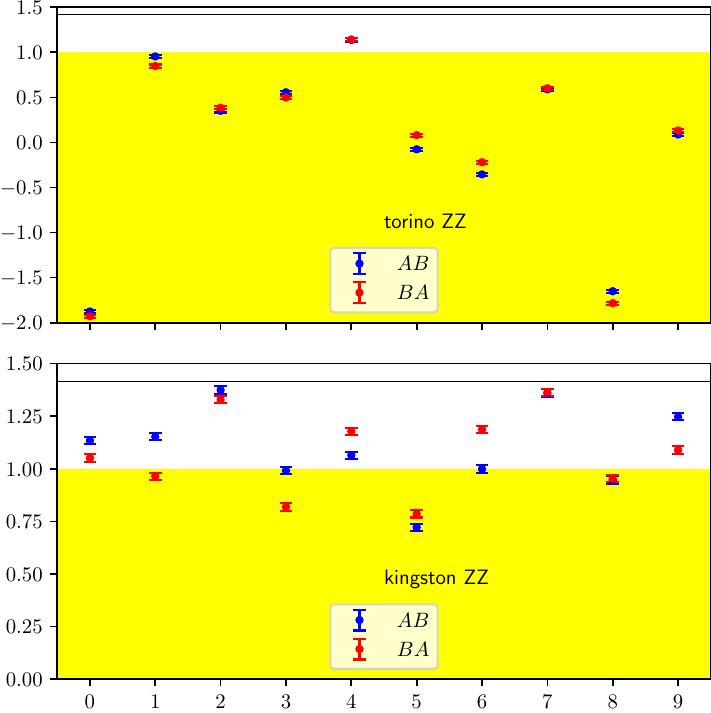}
		\caption{Leggett-Garg violation for fractional $RZZ$ gate devices  (\texttt{ibm\_torino} and \texttt{ibm\_kingston}) for each qubits set (see Table \ref{fffh}).
			The values correspond to the combination of correlations taken from Eq. (\ref{lgab}), while $AB$ and $BA$ correspond to the order $A\to B$ and $B\to A$.
			The solid line is the perfect value $\sqrt{2}$ while the yellow region is below the classical limit.
		}
		\label{lgt}
	\end{figure}

\begin{figure}[ht]
		\includegraphics[scale=.5]{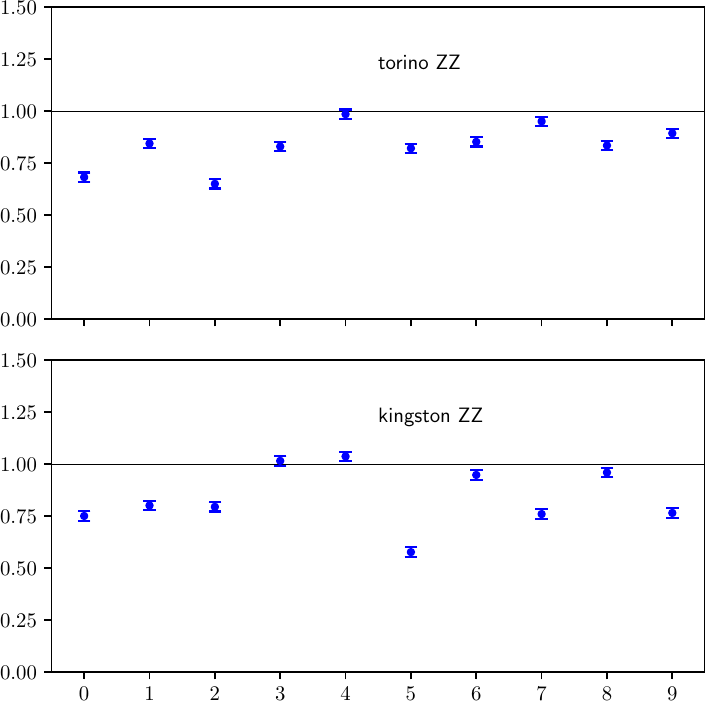}
		\caption{Order invariance violation for fractional $RZZ$ gate devices  (\texttt{ibm\_torino} and \texttt{ibm\_kingston}) for each qubits set (see Table
			\ref{fffh}).
			The values correspond to the difference $\langle \mathcal A\mathcal B\mathcal C\rangle-\langle\mathcal B\mathcal A\mathcal C\rangle$.
			The solid line is the perfect value $1$.
		}
		\label{ordt}
	\end{figure}

	The results were different on the \texttt{ibm\_torino} and \texttt{ibm\_kingston}, with fractional $(ZZ)_\theta$ gate. 
	We collected the data using the same $\theta$ but with $Y_{\pm}$ in the weak measurement circuit, see Fig. \ref{weakf}. This way we were able to get the 	
	measurement strength parametrized by $\pm\theta$. LG inequality was violated only for qubit group $4$ on \texttt{ibm\_torino} but
	at least 5 groups on \texttt{ibm\_kingston}, see Fig. \ref{lgt}. However, all
	the groups passed the test of time order, see \ref{ordt}. Surprisingly, only the
	difference between correlations for $ABC$ and $BAC$ orders remains roughly stable,
	while individual correlations vary across qubit groups. 
	
	We have also tested IonQ, Forte Enterprise 1, on a single set of 3 qubits. 
	Formally those were $0,1,2$, but ion trap technology makes them all equivalent. In those
	experiments we ran 2 jobs, 100000 shots and 25 repetitions each. The circuit was 
	formally identical to the one ran on \texttt{ibm\_torino} and \texttt{ibm\_kingston}. The tested device contains 36
	ions ${}^{171}\mathrm{Yb}^+$, with the drive frequency 12.64GHz between hyperfine
	levels, in a linear Paul trap with expected average 1 and 2--qubit gate
	errors $0.08\%$ and $5.14\%$, respectively. For technical details see the documentation
	\cite{iondoc,avsion,wright_phd,egan_phd,debnath_phd,landsman_phd,figgatt_phd}.
	
	The results, presented in Table \ref{tabi}, show violation of LG inequality and order
	dependence, even exceeding quantum expectations. The origin of that phenomenon could be
	a miscalibration of $(ZZ)_\theta$ gate for the angle $\theta$ passed to the device. 
	A deviation by about $5\%$ from the actual $\theta$, i.e. of the order of 2-qubit gate
	error, would explain the excess results. Due to longer gate times and repetition rates,
	the whole test took about two months, comparing to several hours on IBM.
	
	All correlations and averages are presented in Methods \ref{appe}. The data and the 
	scripts we used are available publicly \cite{zen}.
	
	\begin{table}
		\begin{tabular}{*{3}{c}}
			quantity &value&error\\
			\midrule
			LG$AB$&1.583 & 0.025\\
			LG$BA$&1.569 & 0.025\\
			order&1.190 & 0.032\\
			$\langle ABC\rangle$&0.630& 0.022\\
			$\langle BAC\rangle$&-0.560& 0.022\\
			$\langle AB\rangle$&0.020&0.022\\
			$\langle BA\rangle$&0.004&0.022\\
			$\langle AbC\rangle$&-0.768&0.002\\
			$\langle bAC\rangle$&-0.786&0.002\\
			$\langle aBC\rangle$&0.790&0.002\\
			$\langle BaC\rangle$&0.783&0.002\\
			$\langle Ab\rangle$&0.799&0.002\\
			$\langle bA\rangle$&0.787&0.002\\
			$\langle aB\rangle$&0.790&0.002\\
			$\langle Ba\rangle$&0.800&0.002\\
			$\langle abC\rangle$&0.011&0.0002\\
			$\langle baC\rangle$&0.008&0.0002\\
			\bottomrule
		\end{tabular}
		\caption{The values and errors from the test on IonQ. Here LG$AB$ and LG$BA$ refer
			to LG inequalities Eq. (\ref{lgab}) for the order $A\to B$ and $B\to A$,
			and order = $\langle ABC\rangle-\langle BAC\rangle$. The remaining correlations and
			averages follow the rule: capital letter $A$, $B$ -- outcome matters, small letter
			$a$, $b$ outcome ignored. The order of letters corresponds to the order of
			measurements. For instance $\langle AbC\rangle$ means correlation between $A$ and
			$C$ with the weak measurement of $B$, which occurred after $A$, but was ignored.}
		\label{tabi}
	\end{table}

	\section{Discussion}
	We have shown that weak measurements can be realized on IBM Quantum and similar public
	quantum computers to demonstrate incompatibility between noninvasiveness and realism, by
	means of LG and time order violation. In the case of LG, we had to assume the strength
	of the measurement, so further improvements on LG test are needed in this respect.
	Both violations are significant. One can have an objection that the gates are strong
	and invasive, even in the case of self-reversal. It might be possible to replace the native $ECR$ gate
	by its customized (via pulse modification) version, reducing the interaction
	strength. The newly introduced fractional single-qubit $RX$ gate works excellent on IBM Heron devices, however its two-qubit relative, $RZZ$, works significantly worse. The question of whether the strength of fractional gates is accurately reduced remains ambiguous. On one hand, the time order violation is
	clear, on the other hand, the LG inequality is randomly violated. It seems that IonQ Forte and \texttt{ibm\_kingston} are ahead,
	proving the technological progress in both implementations. Nevertheless, large deviations from theoretical expectations, despite small
	declared error, need urgent explanation. Another open question concerns usefulness of weak measurements in quantum
	computation. Although the price for the small invasiveness is the necessity of large
	statistics, the actual trade-off is yet to be evaluated.

	\begin{figure}[ht]
		\includegraphics[scale=.5]{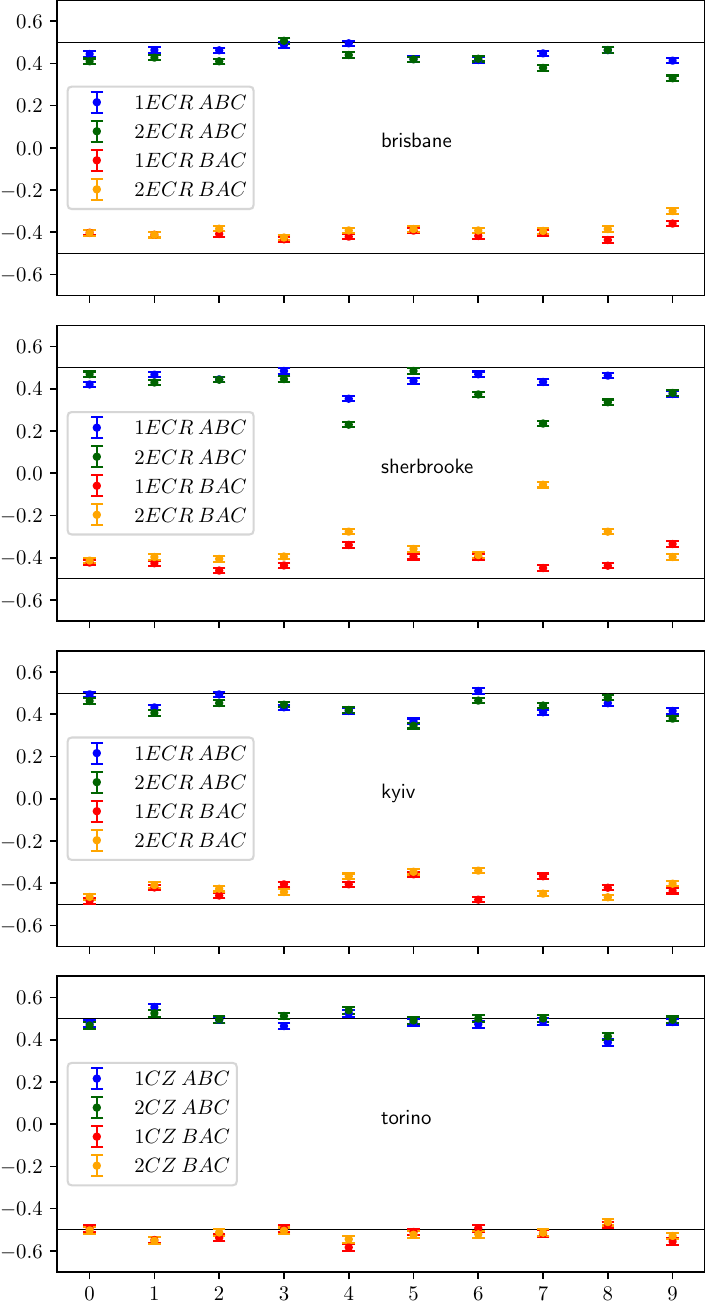}
		\caption{Correlations $\langle \mathcal A\mathcal B\mathcal C\rangle$ and
			$\langle \mathcal B\mathcal A\mathcal C\rangle $
			on he respective IBM devices. Description $nECR/CZ$ corresponds to
			$n=1,2$ $ECR/CZ$ gate(s) in the protocol, while $ABC$ and $BAC$ correspond to the
			order $A\to B$ and $B\to A$. The solid lines are the perfect values $\pm 0.5$.
		}
		\label{ABC}
	\end{figure}
	
	\begin{figure}[ht]
		\includegraphics[scale=.5]{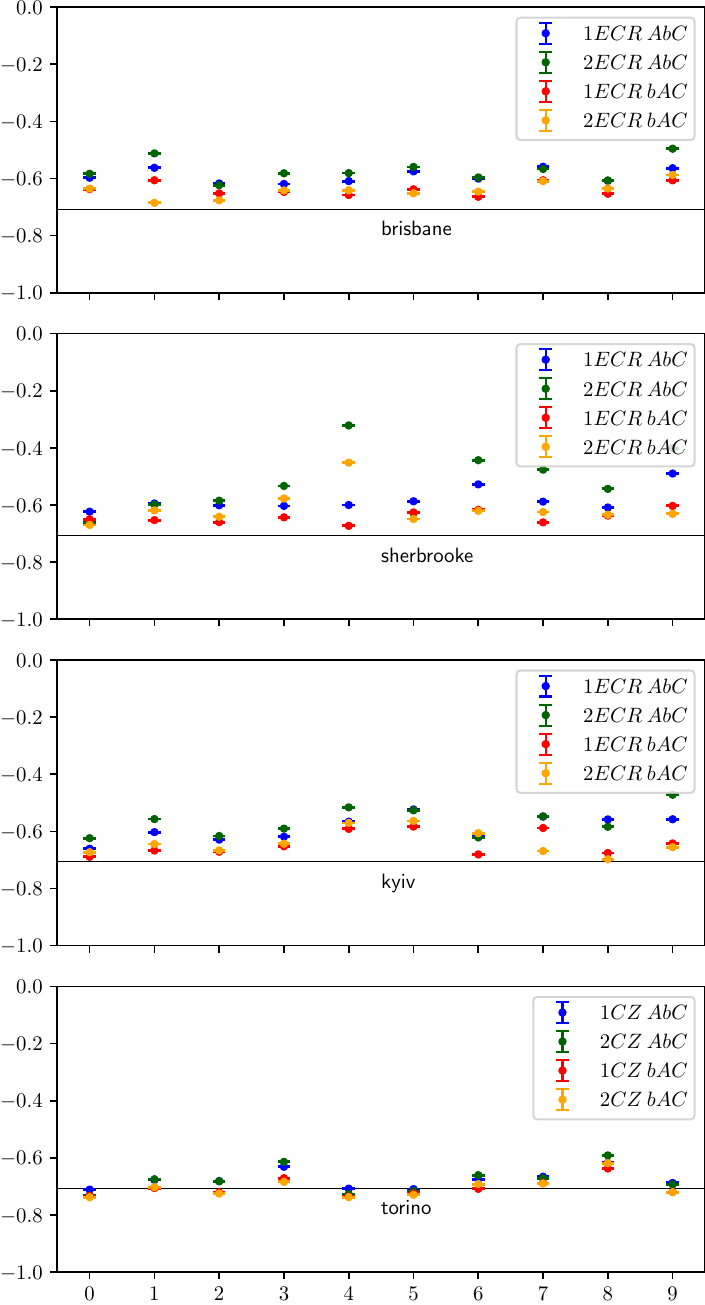}
		\caption{Correlations $\langle \mathcal A\mathcal C\rangle$ on the respective
			IBM devices. Description $nECR/CZ$ corresponds to
			$n=1,2$ $ECR/CZ$ gate(s) in the protocol, while
			$AbC$ and $bAC$ correspond to the order $A\to B$ and $B\to A$. The solid line is
			the perfect value $-1/\sqrt{2}$.
		}
		\label{AC}
	\end{figure}
	
	\begin{figure}[ht]
		\includegraphics[scale=.5]{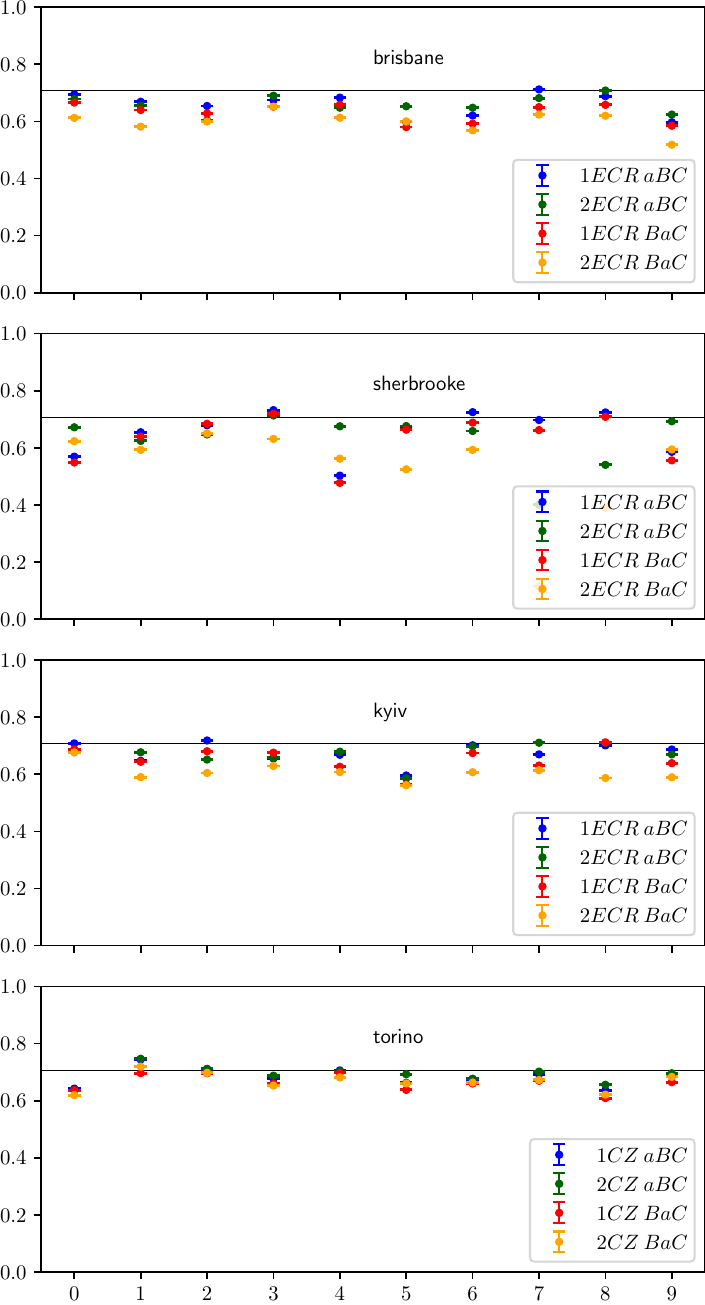}
		\caption{Correlations $\langle \mathcal B\mathcal C\rangle$ on the respective
			IBM devices. Description $nECR/CZ$ corresponds to
			$x=1,2$ $ECR/CZ$ gate(s) in the protocol, while
			$aBC$ and $BaC$ correspond to the order $A\to B$ and $B\to A$. 	The solid lines are
			the perfect values $1/\sqrt{2}$.
		}
		\label{BC}
	\end{figure}
	
	\begin{figure}[ht]
		\includegraphics[scale=.5]{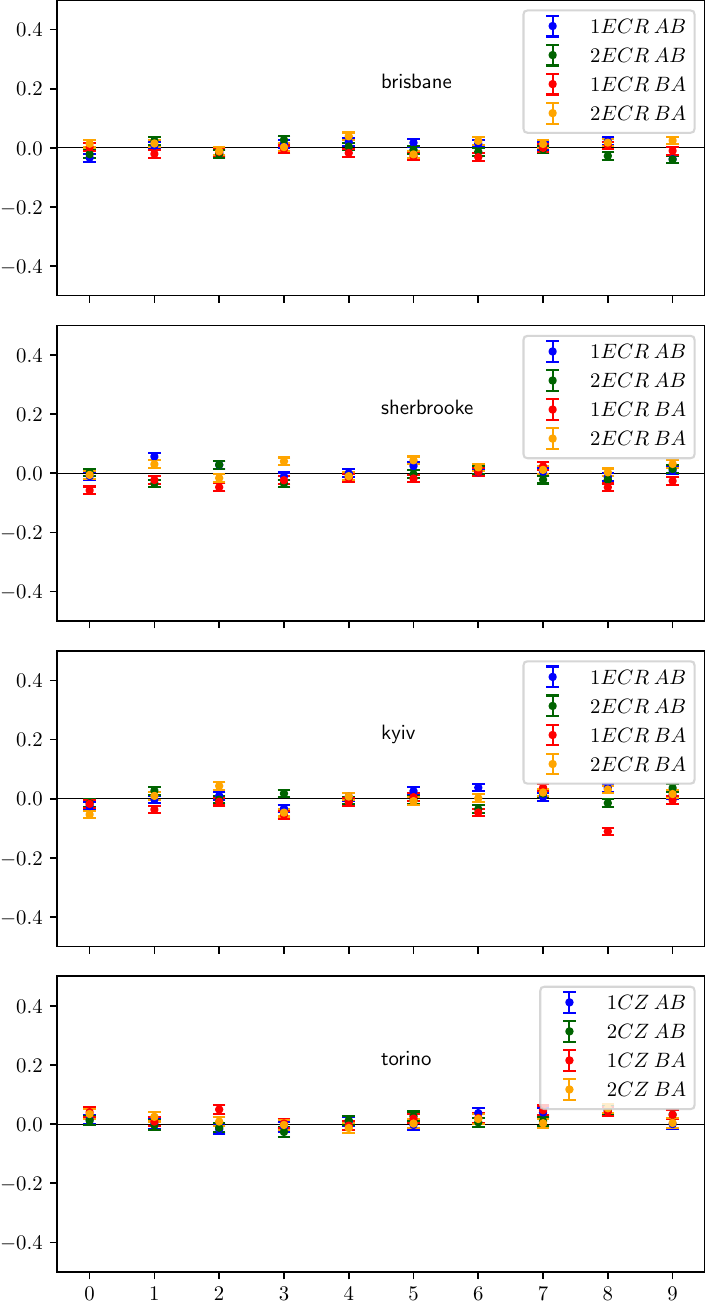}
		\caption{Correlations $\langle \mathcal A\mathcal B\rangle$ and
			$\langle \mathcal B\mathcal A\rangle $ on the respective IBM devices.
			Description $nECR/CZ$ corresponds to
			$n=1,2$ $ECR/CZ$ gate(s) in the protocol, while $AB$ and $BA$
			correspond to the order $A\to B$ and $B\to A$. The solid lines are the perfect
			values $0$.
		}
		\label{AB}
	\end{figure}
	
	\begin{figure}[ht]
		\includegraphics[scale=.5]{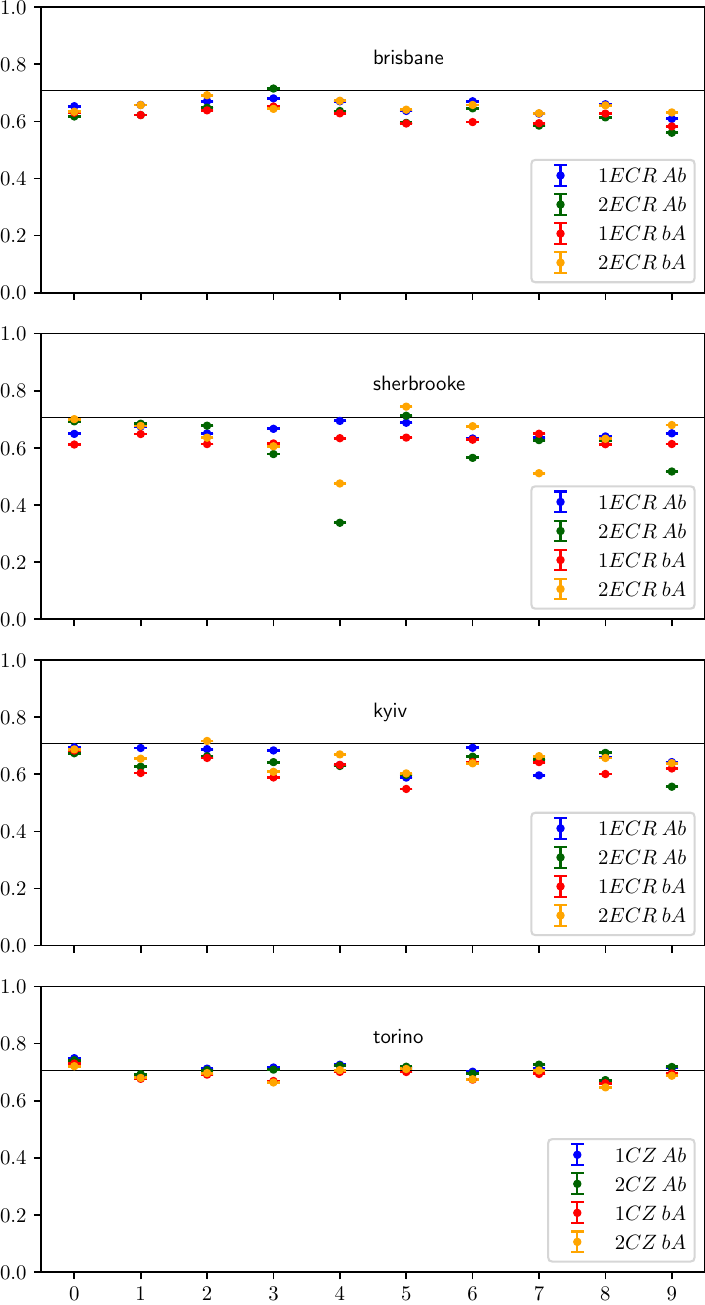}
		\caption{Averages $\langle \mathcal A\rangle$ on the respective IBM devices.
			Description $nECR/CZ$ corresponds to
			$n=1,2$ $ECR/CZ$ gate(s) in the protocol, while $Ab$ and $bA$
			correspond to the order $A\to B$ and $B\to A$. The solid lines is the perfect
			value $1/\sqrt{2}$.
		}
		\label{A}
	\end{figure}
	
	\begin{figure}[ht]
		\includegraphics[scale=.5]{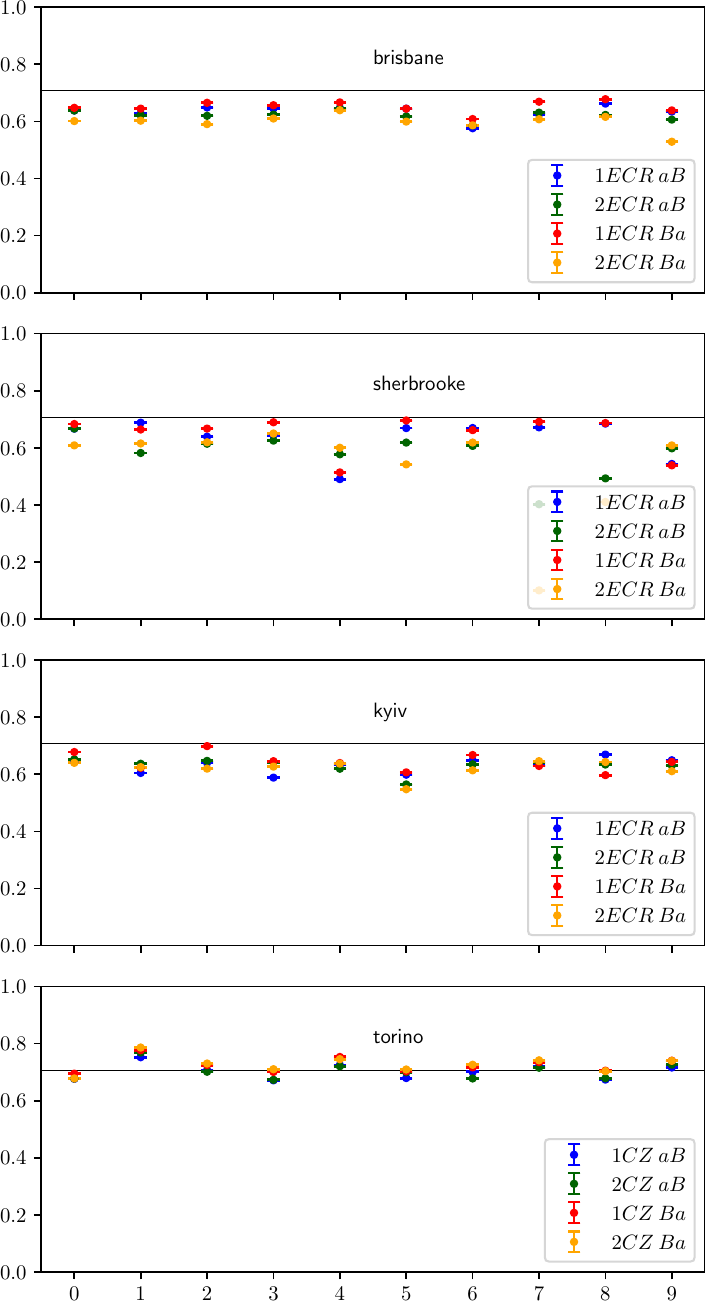}
		\caption{Averages $\langle \mathcal B\rangle$ on the respective IBM devices.
			Description $nECR/CZ$ corresponds to
			$n=1,2$ $ECR/CZ$ gate(s) in the protocol, while $aB$ and $Ba$
			correspond to the order $A\to B$ and $B\to A$. The solid lines is the perfect
			value $1/\sqrt{2}$.
		}
		\label{B}
	\end{figure}
	
	\begin{figure}[ht]
		\includegraphics[scale=.5]{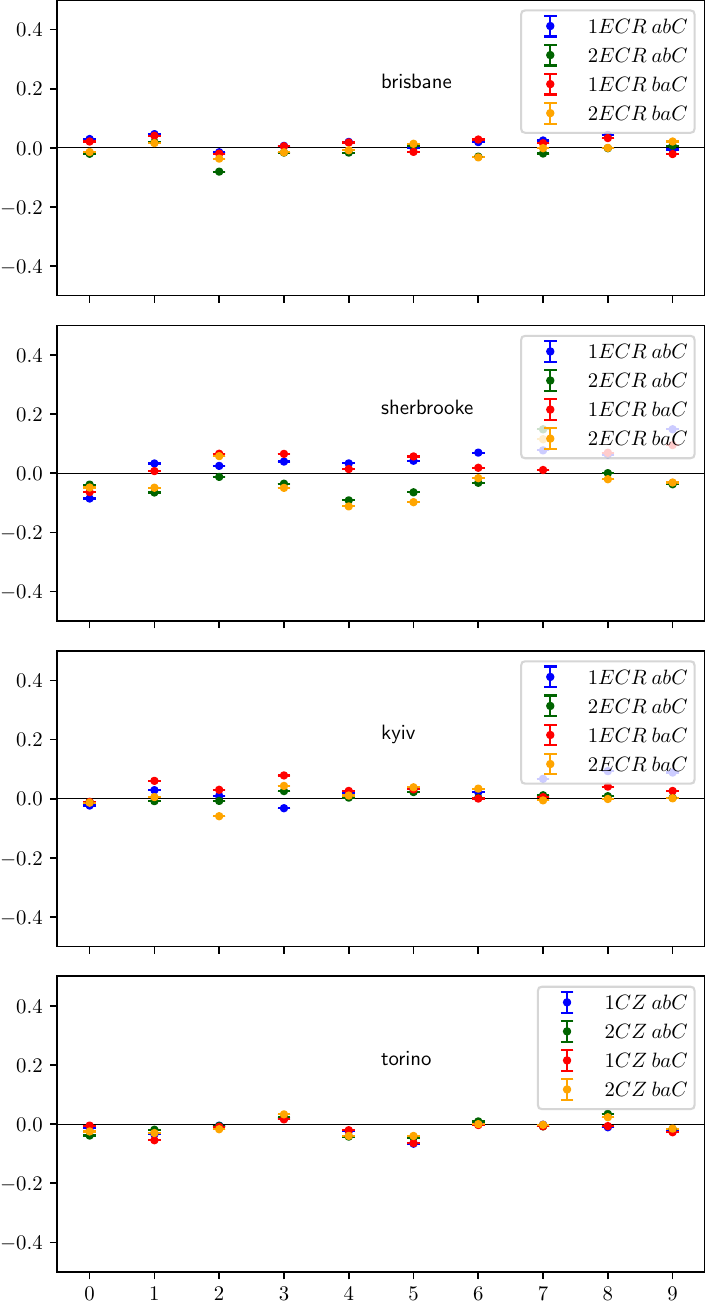}
		\caption{Averages $\langle \mathcal C\rangle$ on the respective IBM devices.
			Description $nECR/CZ$ corresponds to
			$n=1,2$ $ECR/CZ$ gate(s) in the protocol, while $ab$ and $ba$
			correspond to the order $A\to B$ and $B\to A$. The solid line is the perfect
			value $0$.
		}
		\label{C}
	\end{figure}
	
	\begin{figure}[ht]
		\includegraphics[scale=.5]{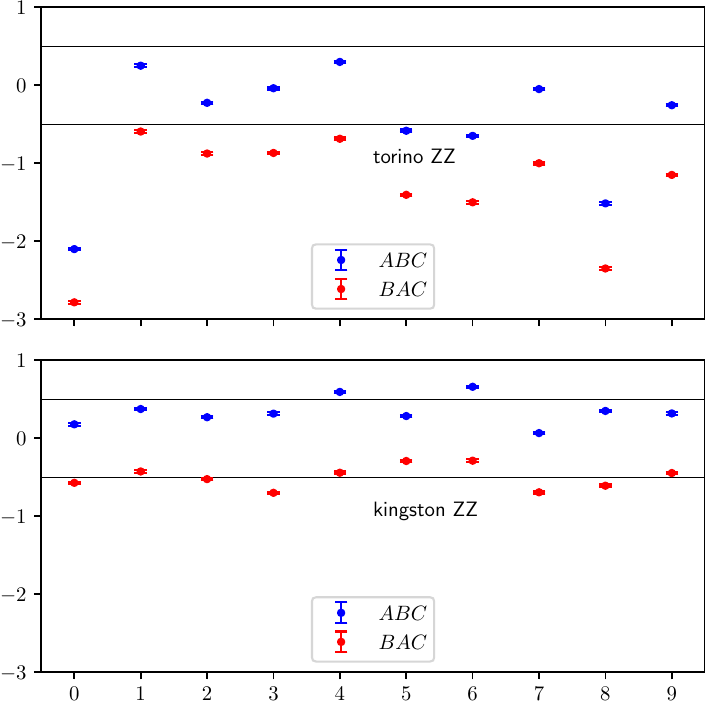}
		\caption{Correlations $\langle \mathcal A\mathcal B\mathcal C\rangle$ and $\langle
			\mathcal B\mathcal A\mathcal C\rangle$ \texttt{ibm\_torino} and \texttt{ibm\_kingston} with $RZZ$ gates. Here $AB$ and $BA$
			correspond to the order $A\to B$ and $B\to A$. The solid lines are the perfect
			values $\pm 0.5$.
		}
		\label{ABCt}
	\end{figure}
	
	\begin{figure}[ht]
		\includegraphics[scale=.5]{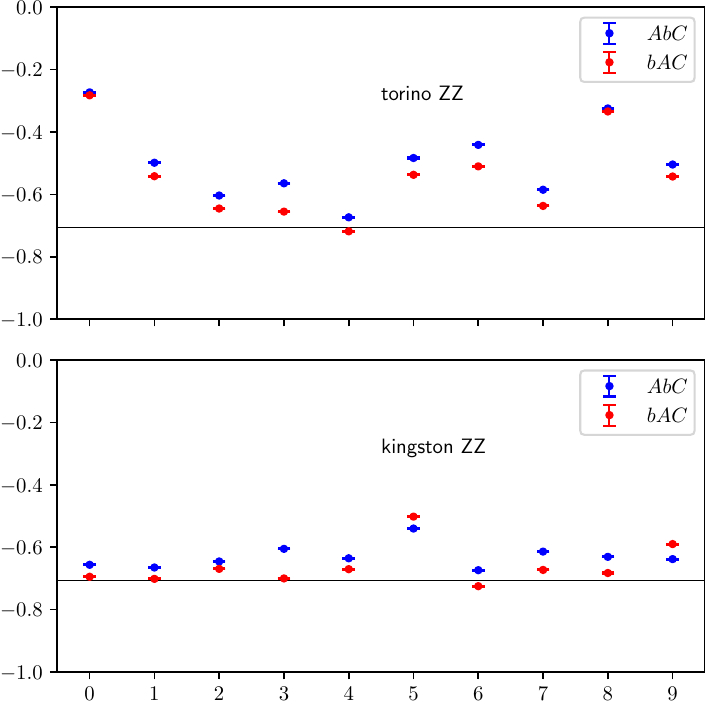}
		\caption{Correlations $\langle \mathcal A\mathcal C\rangle$ \texttt{ibm\_torino} and \texttt{ibm\_kingston} with $RZZ$ gates.
			Here $AbC$ and $bAC$ correspond to the order $A\to B$ and $B\to A$.
			The solid line is the perfect value $-1/\sqrt{2}$.
		}
		\label{ACt}
	\end{figure}
	
	\begin{figure}[ht]
		\includegraphics[scale=.5]{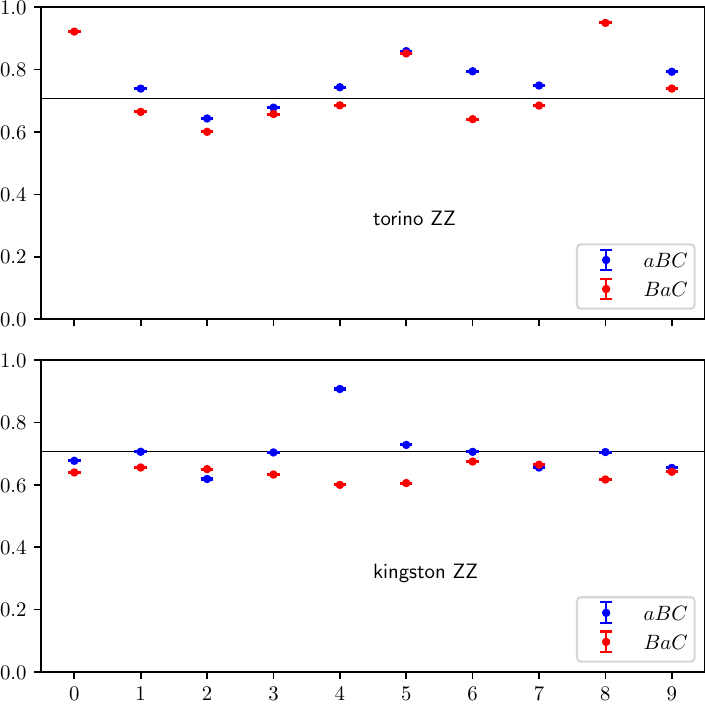}
		\caption{Correlations $\langle \mathcal B\mathcal C\rangle$ \texttt{ibm\_torino} and \texttt{ibm\_kingston} with $RZZ$ gates.
			Here $aBC$ and $BaC$ correspond to the order $A\to B$ and $B\to A$.
			The solid lines are the perfect values $1/\sqrt{2}$.
		}
		\label{BCt}
	\end{figure}
	
	\begin{figure}[ht]
		\includegraphics[scale=.5]{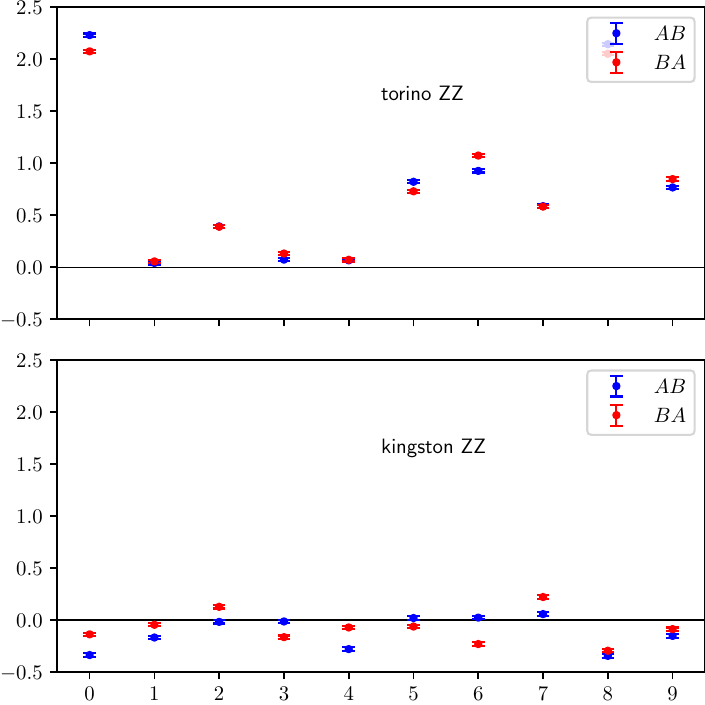}
		\caption{Correlations $\langle \mathcal A\mathcal B\rangle$ and
			$\langle \mathcal B\mathcal A\rangle $ on \texttt{ibm\_torino} and \texttt{ibm\_kingston} with $RZZ$ gates.
			Here $AB$ and $BA$ correspond to the order $A\to B$ and $B\to A$.
			The solid lines are the perfect values $0$.
		}
		\label{ABt}
	\end{figure}
	
	\begin{figure}[ht]
		\includegraphics[scale=.5]{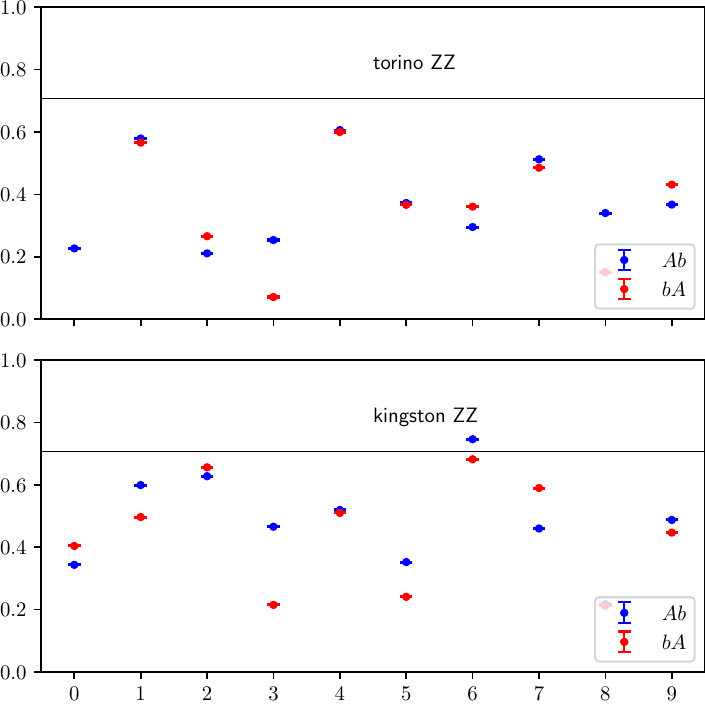}
		\caption{Averages $\langle \mathcal A\rangle$ on \texttt{ibm\_torino} and \texttt{ibm\_kingston} with $RZZ$ gates.
			Here $Ab$ and $bA$ correspond to the order $A\to B$ and $B\to A$.
			The solid lines is the perfect value $1/\sqrt{2}$.
		}
		\label{At}
	\end{figure}
	
	\begin{figure}[ht]
		\includegraphics[scale=.5]{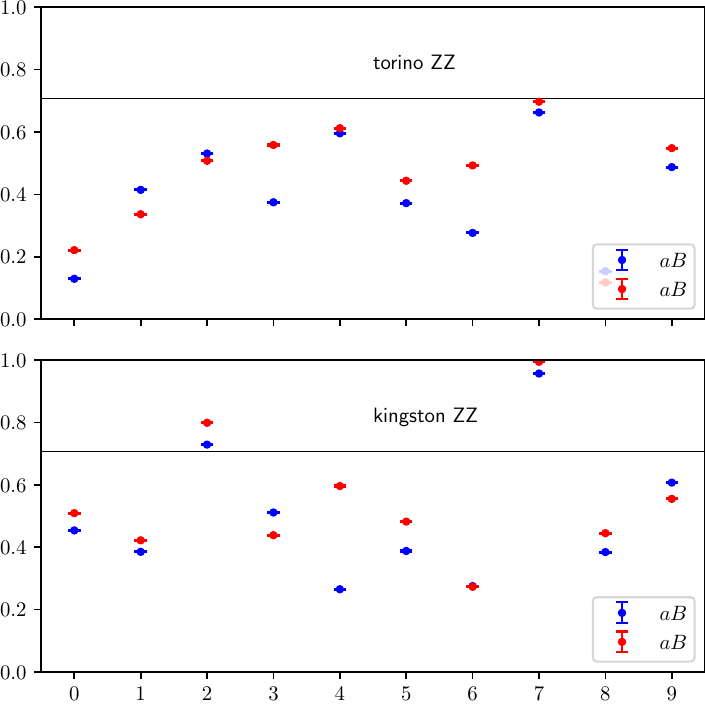}
		\caption{Averages $\langle \mathcal B\rangle$ on \texttt{ibm\_torino} and \texttt{ibm\_kingston} with $RZZ$ gates.
			Here $aB$ and $Ba$ correspond to the order $A\to B$ and $B\to A$.
			The solid lines is the perfect value $1/\sqrt{2}$.
		}
		\label{Bt}
	\end{figure}
	
	\begin{figure}[ht]
		\includegraphics[scale=.5]{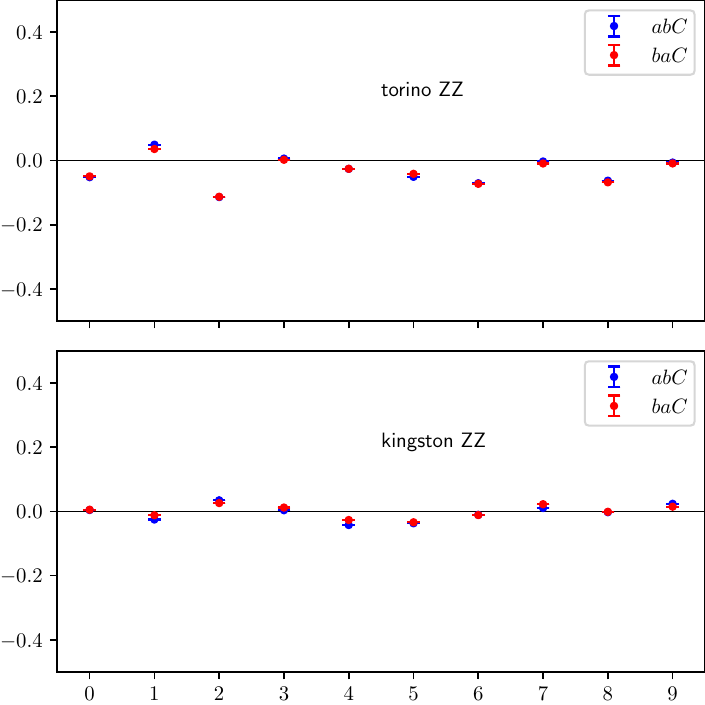}
		\caption{Averages $\langle \mathcal C\rangle$ on \texttt{ibm\_torino} and \texttt{ibm\_kingston} with $RZZ$ gates.
			Here $ab$ and $ba$ correspond to the order $A\to B$ and $B\to A$.
			The solid line is the perfect value $0$.
		}
		\label{Ct}
	\end{figure}

	\section{Methods}
	\subsection{Proof of the Theorem 1}
	\label{appa}
	
	By convexity we can assume that $\rho_\pm=|\pm_0\rangle\langle\pm_0 |$ for some orthogonal states $|\pm_0\rangle$.
	Let us assume $\langle a\rangle_\pm=\pm\Delta/2$ by shifting $a\to a-(\langle a\rangle_{+}+\langle a\rangle_-)/2$
	We define two states, being the superpositions of $|\pm_0\rangle$, i.e.
	\be
	\sqrt{2}|\pm\rangle=|+_0\rangle\pm |-_0\rangle.
	\ee
	Then 
	\begin{align}
		&\Delta=\int da\; a\mathrm{Tr}K^\dag(a)K(a) (|+_0\rangle\langle +_0|-|-_0\rangle\langle -_0|)\nonumber\\
		&=\int da\; a\mathrm{Tr}K^\dag(a)K(a)(|-\rangle\langle +|+|+\rangle\langle -|).
	\end{align}
	We apply the Cauchy-Schwarz inequality $\langle v|v\rangle\langle w|w\rangle\geq |\langle v|w\rangle|^2$
	to 
	\begin{align}
		&|v\rangle= aK(a)|\pm\rangle,\nonumber\\
		&|w\rangle=(P_\mp+P/2)K(a)|\mp\rangle
	\end{align}
	for
	$P_\pm=|\pm\rangle\langle \pm|$ and $P=1-P_+-P_-$ being the projectors onto three relevant subspaces.
	then
	\be
	\langle w|w\rangle=\langle K^\dag(a)(P_\mp+P/4)K(a)\rangle_\mp
	\ee
	and
	\begin{align}
		&\sum_{\pm}\int da\;\sqrt{\langle K^\dag(a)a^2K(a)\rangle_\pm\langle K^\dag(a)(P_\pm+P/4)K(a)\rangle_\mp}\nonumber\\
		&\geq \Delta/2
	\end{align}
	with the subscript $\pm$ for the states $\rho=P_\pm$.
	We can replace $P/4$ by $P$ not decreasing the left hand side.
	By the Cauchy-Schwarz inequality
	\be
	\left(\int fgda\right)^2\leq \int f^2da\int g^2da
	\ee
	applied to
	\begin{align}
		&f^2=\langle K^\dag(a)a^2K(a)\rangle_\pm,\nonumber\\
		&g^2=\langle K^\dag(a)(P_\pm+P)K(a)\rangle_\mp
	\end{align}
	we get
	\be
	\sum_{\pm}\sqrt{\langle a^2\rangle_\mp \eta_\pm}\geq \Delta/2
	\ee
	for
	\be
	\eta_\pm=1-\int da\;|\langle K(a)\rangle_\pm|^2
	\ee
	We chose $\rho=P_\pm$ with the larger value $\langle a^2\rangle_\mp \eta_\pm$
	and finally get
	\be
	16\sigma^2E\geq\Delta^2
	\ee
	which completes the proof.
	
	\subsection{Measurements of dichotomic observables with arbitrary strength}
	\label{appb}
	
	The full treatment of weak measurement of the strength $\lambda$ requires two superoperators
	\be
	\mathcal A\rho=\sum_a aK_\lambda(a)\rho K^\dag_\lambda(a)/\lambda
	\ee
	with 
	\be
	K_\lambda(a)=\sqrt{(1+\lambda aA)/2}
	\ee
	and $a=\pm 1$, i.e. outcomes of the measurement on the meter qubits,
	which reduces to
	\be
	\sqrt{2}K_\lambda(a)=\cos(\theta/2)+A\sin(\theta/2)
	\ee
	and $\mathcal A=\{A,\cdot\}/2$ for $\lambda=\sin\theta$.
	If one ignores the measurment, there is still invasiveness
	\be
	\tilde{\mathcal A}\rho=\sum_a K_\lambda(a)\rho K^\dag_\lambda(a)
	\ee
	reducing to
	\be
	\tilde{A}=1-g\mathcal G_A
	\ee
	with $\mathcal G_A=[A,[A,\cdot]]$
	for dichotomic $A$ and
	\be
	g=\sin^2(\theta/2)/2=(1-\sqrt{1-\lambda^2})/4\sim\lambda^2/8
	\ee
	In the eigenbasis $|+\rangle$, $|-\rangle$ of $A=|+\rangle\langle +|-|-\rangle\langle -|$, the superoperators read
	\be
	\mathcal A
	\begin{pmatrix}
		\rho_{++}&\rho_{+-}\\
		\rho_{-+}&\rho_{--}
	\end{pmatrix}
	=\begin{pmatrix}
		\rho_{++}&0\\
		0&\rho_{--}
	\end{pmatrix}
	\ee
	while
	\be
	\tilde{\mathcal A}
	\begin{pmatrix}
		\rho_{++}&\rho_{+-}\\
		\rho_{-+}&\rho_{--}
	\end{pmatrix}
	=\begin{pmatrix}
		\rho_{++}&\rho_{+-}\cos\theta\\
		\rho_{+-}\cos\theta&\rho_{--}
	\end{pmatrix}.
	\ee
	This POVM also captures projection at $\lambda=1$, $\theta=\pi/2$ and $g=1/4$.
	
	Let all observables $A$, $B$, $C$, be dichotomic.
	Note that the strength of the last measurements is irrelevant.
	In the case of a single weak measurement of $A$ followed by  $C$ the readout correlation is equal
	and the backaction corrected measurement read
	\be
	\langle \tilde{\mathcal A}C\rangle=\langle \mathcal C\rangle -\langle \mathcal G_A\mathcal C\rangle g_A,
	\ee
	For two weak ones, first $A\to B$ followed by $C$, we have
	\begin{align}
		&\langle \mathcal A\tilde{\mathcal B}C\rangle=\langle\mathcal A\mathcal C\rangle-\langle \mathcal A\mathcal G_B\mathcal C\rangle g_B,\nonumber\\
		&\langle\tilde{\mathcal A} \mathcal B\mathcal C\rangle=\langle\mathcal B\mathcal C\rangle-\langle \mathcal G_A\mathcal B\mathcal C\rangle g_A,\nonumber\\
		&\langle\tilde{\mathcal A}\tilde{\mathcal B}\mathcal C\rangle=\langle\mathcal C\rangle-\langle (g_A\mathcal G_A+g_B\mathcal G_B)\mathcal C\rangle
		+\langle \mathcal G_A\mathcal G_B\mathcal C\rangle g_Ag_B,
	\end{align}
	Here $\tilde{\mathcal O}$ means that $O$ is measured with ignored outcomes, so it may affect other obsrevables.
	
	\subsection{IBM Quantum circuits}
	\label{appc}

	The default measured observable  $Z$, can be changed $Z\to U^\dag Z U$ by rotation $U$ on the qubit,
	\be
	X=Y_+ZY_-,\:Y=X_-ZX_+
	\ee
	expressed by native gates  $Y_\pm=Z_\pm X_+ Z_\mp$ and $X_-=ZX_+Z$. We shall also use $X$ rotated by $\pm \pi/4$ about $Z$ axis, i.e.
	\be
	X^\pm=Z_{\pm\pi/4}X Z_{\mp\pi/4}=(X\pm Y)/\sqrt{2}
	\ee
	to get $A=X^-$, $B=X^+$, $C=Z$. The initial state is obtained by $|\psi\rangle=Y_+|0\rangle$.
	Then $[A,B]=[X^-,X^+]=iZ$ and $[Z,Y]=-iX$ but $\rho=|\psi\rangle\langle \psi|$ is its eigenstate,
	so 
	\begin{align}
		&\langle \mathcal C\rangle =0,\nonumber\\
		&\langle \mathcal A \rangle=\langle\mathcal B\rangle=1/\sqrt{2},\nonumber\\
		&\langle \mathcal A\mathcal B\rangle= \langle \mathcal B\mathcal A\rangle=0,\nonumber\\
		&\langle \mathcal B\mathcal C\rangle=1/\sqrt{2}=-\langle \mathcal A\mathcal C\rangle,\nonumber\\
		&\langle \mathcal A\mathcal B\mathcal C\rangle=1/2=-\langle \mathcal B\mathcal A\mathcal C\rangle
	\end{align}
	Now the LG inequalities should be violated
	\be
	\langle\mathcal A\rangle+\langle\mathcal B\rangle-\langle\mathcal A\mathcal B\rangle=\sqrt{2}\label{lgab}
	\ee
	but also
	\begin{align}
		&\langle\mathcal A\rangle+\langle\mathcal C\rangle-\langle\mathcal A\mathcal C\rangle=\sqrt{2},\nonumber\\
		&\langle\mathcal B\rangle-\langle\mathcal C\rangle+\langle\mathcal B\mathcal C\rangle=\sqrt{2},
	\end{align}
	which can be reduced to (\ref{lgg}) replacing $C$ by $B$ or $-A$.
	
	\subsection{Corrections from finite strength}
	\label{appd}
	
	To find corrections of finite invasiveness to the measurement $A\to B\to C$, we note that
	$\mathcal G_BC=-2\sqrt{2}A$, $\mathcal G_AC=2\sqrt{2}B$, $\mathcal G_AB=4B$, so
	\begin{align}
		&\langle \mathcal A\tilde{\mathcal B}\mathcal C\rangle=-1/\sqrt{2}+2\sqrt{2}g_B,\nonumber\\
		&\langle \tilde{\mathcal A}\mathcal B\mathcal C\rangle=1/\sqrt{2},\nonumber\\
		&\langle \tilde{\mathcal A}\tilde{\mathcal B} C\rangle=2(g_B-g_A),\nonumber\\
		&\langle \tilde{\mathcal A}\mathcal B\rangle=1/\sqrt{2}-2\sqrt{2}g_A
	\end{align}
	For $B\to A\to C$ we get $\mathcal G_BA=4A$, and so
	\begin{align}
		&\langle \mathcal B\tilde{\mathcal A}\mathcal C\rangle=1/\sqrt{2}-2\sqrt{2}g_A,\nonumber\\
		&\langle \tilde{\mathcal B}\mathcal A\mathcal C\rangle=-1/\sqrt{2},\nonumber\\
		&\langle \tilde{\mathcal B}\tilde{\mathcal A}\mathcal C\rangle=2(g_B-g_A),\nonumber\\
		&\langle \tilde{\mathcal B}\mathcal A\rangle=1/\sqrt{2}-2\sqrt{2}g_B.
	\end{align}
	Strong measurements $g=1/4$ end up with highly disturbed results
	for $A\to B\to C$,
	\be
	\langle  \mathcal A\tilde{\mathcal B}\mathcal C\rangle=\langle \tilde{\mathcal A}\mathcal B\rangle=0
	\ee
	and for $B\to A\to C$,
	\be
	\langle \mathcal B\tilde{\mathcal A}\mathcal C\rangle=\langle \tilde{\mathcal B}\mathcal A\rangle=0
	\ee
	The special case is $\langle \tilde{\mathcal B}\tilde{\mathcal A}\mathcal C\rangle=\langle \tilde{\mathcal A}\tilde{\mathcal B} C\rangle$
	which remain $0$ for equal strength but bounds $1/2$ projection on $A$ and weak measurement of $B$ and $-1/2$ vice versa.

	\subsection{Full results of the test}
	\label{appe}
	
	Our tests allow to extract all averages and correlations, presented in Figs. \ref{ABC}, \ref{AB}, \ref{AC}, \ref{BC}, \ref{A}, \ref{B}, \ref{C},
	and Figs. \ref{ABCt}, \ref{ABt}, \ref{ACt}, \ref{BCt}, \ref{At}, \ref{Bt}, \ref{Ct}.
	These results, except $\langle \mathcal A\mathcal B\rangle$, $\langle \mathcal A\mathcal B\mathcal C\rangle$ and $A\leftrightarrow B$, 
	differ clearly from projective measurements, confirming again the weak measurement regime. They roughly agree with the theoretical predictions.

		\section*{Acknowledgments}
	
	The results have been created using IBM Quantum. The views expressed are those of the
	authors and do not reflect the official policy or position of the IBM Quantum team.
	We thank Poznań Supercomputing and Networking Center and ICM UW for the access to IBM Quantum Innovation Center.

	\section*{Abbreviations}
ECR, echoed cross-resonance.
\section*{Declarations}
\subsection*{Ethical Approval and Consent to participate}
Not applicable.
\subsection*{Consent for publication}
Not applicable.
\subsection*{Availability of supporting data}
The data are publicly available at
	doi.org/10.5281/zenodo.16402161
\subsection*{Competing interests}
All the authors declare no competing interests.
\subsection*{Funding}
	This work was supported in part by program "{}Excellence
	initiative -- research university"{} for the AGH University in Krakow as well as the
	ARTIQ project: UMO-2021/01/2/ST6/00004 and ARTIQ/0004/2021. 
\subsection*{Authors' contributions}
T.R. developed the idea and simulations, T.B. and T.R.  wrote codes,  J.B. and A.B. wrote the main manuscript text, B.Z. and A.S. helped with resources to run experiments, W.B. supervised the manuscript preparation and improved the presentation.  All authors reviewed the manuscript.

\end{document}